\documentstyle[11pt,bnl,epsf]{article}

\begin{document}

\title{{\bf The Effective Electroweak Chiral Lagrangian:\\
The Matter Sector.}}
\author{{{\sc E. Bagan}\thanks{%
bagan@ifae.es}} \\
{Grup de F\'\i sica Te\`orica and IFAE}\\
{Universitat Aut\`onoma de Barcelona,
E-08193 Bellaterra}\\
\\
{{\sc D. Espriu}\thanks{%
espriu@greta.ecm.ub.es}}\ and \ {{\sc J. Manzano}\thanks{%
manzano@ecm.ub.es}} \\
{Departament d'Estructura i Constituents de la Mat\`eria and IFAE} \\
{\ Universitat de Barcelona,
 Diagonal, 647,
E-08028 Barcelona}}
\date{}
\maketitle
\begin{abstract}
We parametrize in a model-independent way possible departures from the
minimal Standard Model predictions in the matter sector.
We only assume the symmetry breaking pattern of the Standard Model and
that new particles are sufficiently heavy so that the symmetry is
non-linearly realized. Models with dynamical symmetry breaking
are generically of this type.  We review in the
effective theory language to what extent the simplest models of dynamical
breaking are actually constrained
and the assumptions going into the comparison with experiment.
Dynamical
symmetry breaking models can be approximated at intermediate energies by
four-fermion
operators. We present a complete classification of the latter
when new particles appear in the usual representations of
the $SU(2)_L\times SU(3)_c$ group as well as a partial
classification  in the general case.
We discuss the accuracy of the four-fermion description
by matching to a simple `fundamental' theory. The coefficients
of the effective lagrangian in the matter sector for
dynamical symmetry breaking models (expressed in terms of the
coefficients of the four-quark operators) are then compared
to those of models with elementary scalars (such as the minimal Standard
Model).
Contrary to a somewhat widespread belief, we see that
the sign of the vertex corrections is not fixed in dynamical symmetry
breaking models. This work provides the theoretical tools required to
analyze, in a rather general setting, constraints on the
matter sector of the Standard Model.

\end{abstract}

\vfill
\vbox{
UB-ECM-PF 98/17\null\par}

\clearpage

\section{Introduction}

The Standard Model of electroweak interactions has by now been
impressively tested to one part in a thousand level thanks to the formidable
experimental work carried out at LEP and SLC. However,
when it comes to the symmetry breaking mechanism clouds remain
in this otherwise bright horizon.

In the minimal version of the Standard Model of electroweak interactions the
same mechanism (a one-doublet complex scalar field) gives masses
simultaneously to the $W$ and $Z$ gauge bosons and to the fermionic matter
fields (other than the neutrino).
In the simplest minimal Standard Model there is an upper bound
on
$M_H$ dictated by triviality considerations, which hint at the
fact that at a scale $\sim 1$ TeV new interactions should appear if the
Higgs particle is not found by then\cite{triviality}.
On the other hand, in the minimal Standard Model
it is completely unnatural to have a light Higgs particle since its mass is not
protected by any symmetry.

This contradiction is solved by
supersymmetric extensions of the Standard Model, where essentially
the same symmetry breaking mechanism is at work, although the scalar
sector becomes much richer in this case. Relatively light scalars are
preferred. In fact, if supersymmetry is to remain a
useful idea in phenomenology, it is crucial that the Higgs particle is found
with a mass $M_H\le 125$ GeV, or else the theoretical problems, for which
supersymmetry was invoked in the first place, will reappear\cite{hierarchy}.
A very recent two-loop calculation\cite{twoloop} raises this limit somewhat,
to about 130 GeV.

A third possibility is the one provided by models of dynamical symmetry
breaking (such as technicolor (TC) theories\cite{TC}). Here there
are interactions that become strong, typically at the scale $%
\Lambda_\chi\simeq 4\pi v$ ($v=250$ GeV), breaking the global $SU(2)_L\times
SU(2)_R$ symmetry to its diagonal subgroup $SU(2)_V$ and producing Goldstone
bosons which eventually become the longitudinal degrees of freedom of the $%
W^\pm$ and $Z$. In order to transmit this symmetry breaking to the ordinary
matter fields one usually requires additional interactions, characterized by
a different scale $M$. Generally, it is assumed that $M\gg 4\pi v$, to keep
possible flavour-changing neutral currents (FCNC) under
control\cite{FCNC}.
Thus a distinctive
characteristic of these models is that the mechanism giving masses to the $%
W^\pm$ and $Z$ bosons and to the matter fields is different.

Where do we stand at present?
Some will go as far as saying that an elementary Higgs (supersymmetric or
otherwise) has been `seen' through radiative corrections and that its
mass is below 200 GeV. Others dispute this fact (see for instance
\cite{chanowitz} for a critical review of current claims of
a light Higgs).

The effective lagrangian approach has proven remarkably useful in setting
very stringent bounds on some types of new physics taking as input basically
the LEP\cite{aleph} (and SLD\cite{SLD}) experimental results. The idea is to
consider the most general lagrangian which describes the interactions
between the gauge sector and the Goldstone bosons appearing after the $%
SU(2)_L\times SU(2)_R\to SU(2)_V$ breaking takes place. No special mechanism
is assumed for this breaking and thus the procedure is completely general,
assuming of course that particles not explicitly included in the effective
lagrangian are much heavier than those appearing in it. The dependence on
the specific model is contained in the coefficients of higher dimensional
operators. So far only the oblique corrections have been analyzed in this
way.

Our purpose in this work is to extend these techniques to the
matter sector of the Standard Model. We shall write the
leading non-universal operators, determine how their coefficients
affect different physical observables and then determine their value
in two very general families of models: those containing elementary scalars
and those with dynamical symmetry breaking. Since the latter become
non-perturbative at the $M_Z$ scale, effective lagrangian techniques are
called for anyway.  In short, we would like to provide the theoretical tools
required to test ---at least in principle---  whether the mechanism giving
masses to quarks and fermions is the same as that which makes the
intermediate vector
bosons massive or not without having to get involved in
the nitty-gritty details of particular models.
This is mostly
a theoretical paper and we shall leave for a later work a more detailed
comparison with the current data.

\section{The effective lagrangian approach}

Let us start by
briefly recalling the salient features of the effective
lagrangian analysis of the oblique corrections.

Including only those operators
which are relevant for oblique corrections, the effective lagrangian reads
(see e.g. \cite{DEH,eff-lag} for the complete lagrangian)
\begin{equation}
{\cal L}_{\rm eff}=\frac{v^2}4{\rm tr}D_\mu UD^\mu U^{\dagger }+a_0{g^{\prime }}%
^2\frac{v^2}4({\rm tr}TD_\mu UU^{\dagger })^2+a_1gg^{\prime }{\rm tr}UB_{\mu
\nu }U^{\dagger }W^{\mu \nu }-a_8\frac{g^2}{4}({\rm tr}TW^{\mu \nu })^2,
\label{effobl}
\end{equation}
where $U=\exp (\ri \vec{\tau}\cdot \vec{\chi}/v)$ contains the 3 Goldstone
bosons generated after the breaking of the global symmetry $SU(2)_L\times
SU(2)_R\to SU(2)_V$. The covariant derivative is defined by
\begin{equation}
D_\mu U=\p_\mu U+\ri g{\frac{\vec{\tau}}2}\cdot \vec{W}_\mu U-\ri g^{\prime
}U{\frac{\tau _3}2}B_\mu .  \label{30.6a}
\end{equation}
$B_{\mu \nu }$ and $W^{\mu \nu }$ are the
field-strength tensors
corresponding to the right and left gauge groups, respectively
\begin{equation}
W_{\mu\nu}=\frac{\vec\tau}{2} \cdot \vec W_{\mu\nu},\qquad
B_{\mu\nu}=\frac{\tau^3}{2} (\partial_\mu B_\nu-\partial_\nu B_\mu),
\end{equation}
and $T=U\tau
^3U^{\dagger }$. Only terms up to order ${\cal O}(p^4)$ have been included.
The reason is that dimensional counting arguments suppress, at presently
accessible energies, higher dimensional terms, in the hypothesis that
all undetected particles are much heavier than those included in the
effective lagrangian. While the first term on
the r.h.s. of (\ref{effobl}) is universal (in the unitary gauge it is just the
mass term for the $W^{\pm }$ and $Z$ bosons), the coefficients $a_0$, $a_1$
and $a_8$ are non-universal. In other words, they depend on the specific
mechanism responsible for the symmetry breaking. (Throughout this paper the term
`universal' means `independent of the specific mechanism triggering $%
SU(2)_L\times SU(2)_R\to SU(2)_V$ breaking'.)

Most $Z$-physics observables relevant for electroweak physics can be
parametrized
in terms of vector and axial couplings $g_V$ and $g_A$. These are, in
practice, flavour-dependent since they include vertex corrections which
depend on the specific final state. Oblique corrections are however the same
for all final states. The non-universal (but generation-independent)
contributions to $g_V$ and $g_A$ coming from the effective lagrangian (\ref
{effobl}) are
\begin{equation}
\bar g_V=a_0g^{\prime \,2}\left[ I_f^3+2Q_f\left( 2c_W^2-s_W^2\right)
\right] +2a_1Q_fg^2s_W^2+2a_8Q_fg^2c_W^2,  \label{gagvobl}
\end{equation}
\begin{equation}
\bar g_A=a_0I_f^3g^{\prime \,2}\label{gagvvv}.
\end{equation}
They do depend on the
specific underlying breaking mechanism through the values of the $a_i$.
It should be noted that
these coefficients depend logarithmically on some unknown
scale. In the minimal Standard Model the characteristic scale is the Higgs
boson mass, $M_H$. In other theories the scale $M_H$ will be replaced by some other
scale $\Lambda$. A crucial prediction of chiral perturbation theory is
that the dependence on these different scales is logarithmic and actually
the same. It is thus possible
to eliminate this dependence by building suitable combinations of $g_V$ and
$g_A$\cite{EH,EM} determined
by the condition of absence of logs. Whether this line intersects
or not the experimentally allowed region is a direct test of the nature of
the symmetry breaking sector, independently of the precise value of Higgs mass
(in the minimal Standard Model) or of the scale of new interactions (in
other scenarios)\footnote{%
Notice that, contrary to a somewhat widespread belief, the limit $M_H\to \infty$
does not correspond a Standard Model `without the Higgs'. There are some
non-trivial non-decoupling effects}.

One could also try to extract information about the
individual coefficients $a_0$, $a_1$ and $a_8$ themselves, and not only on the
combinations cancelling the dependence on the unknown scale. This
necessarily implies assuming a specific value for the scale $\Lambda$ and
one should be aware that when considering these cut-off dependent quantities
there are finite uncertainties of order $1/16 \pi^2$ associated to
the subtraction procedure
---an unavoidable consequence of using an effective theory, that is often
overlooked. (And recall that using an effective theory is almost mandatory
in dynamical symmetry breaking models.) Only finite combinations of
coefficients have a universal meaning. The subtraction scale uncertainty
persists
when trying to find estimates of the above coefficients via dispersion
relations and the like\cite{PT}.

In the previous analysis it is assumed
that the hypothetical new physics contributions
from vertex corrections are completely negligible. But is it so?
The way to analyze such vertex corrections in a model-independent way is
quite similar to the one outlined for the oblique
corrections. We shall introduce in the next section the most general
effective
lagrangian describing the matter sector. In this sector there is one
universal operator (playing a role analogous to that of the first operator
on the r.h.s. of (\ref{effobl}) in the purely bosonic sector)
\begin{equation}
{\cal L}_{\rm eff}= -v \bar q_{L}U y_{f}q_{R}+ \mbox{h. c.},\qquad y_{f}=y
{\bf 1}%
+y_{3}\tau_{3}.  \label{mass}
\end{equation}
It is an operator of dimension 3. In the unitary gauge $U=1$, it is just the
mass term for the matter fields. For instance if $\bar q_L$ is the doublet $%
(\bar t, \bar b)$
\begin{equation}
m_t=v(y+y_3)=v y_t,\qquad m_b=v(y-y_3)=v y_b.
\end{equation}
Non-universal operators carrying in their coefficients the information on
the mechanism giving masses to leptons and quarks will be of dimension 4 and
higher.

We shall later derive the values of the coefficients corresponding to
operators in the effective lagrangian of dimension 4
within the minimal Standard Model in the large $M_H$ limit and see how the
effective lagrangian provides a convenient way of tracing the Higgs mass
dependence in physical observables. We shall later argue that non-decoupling
effects should be the same
in other theories involving elementary scalars, such as e.g. the two-Higgs
doublet model, replacing $M_H$ by
the appropriate mass.

Large non-decoupling effects appear in theories of dynamical symmetry
breaking and thus they are likely to produce large contributions to the
dimension 4 coefficients. If the scale characteristic of the extended
interactions
(i.e. those responsible of the fermion mass generation) is much larger
than the scale characteristic of the electroweak breaking, it makes
sense to parametrize the former, at least at low energies, via
effective four-fermion operators\footnote{While using an effective theory
description based on four-fermion operators
alone frees us from having to appeal
to any particular model it is obvious that
some information is lost.
This issue turns out to be a rather subtle one
and shall be discussed and quantified in turn. }. We shall assume here that
this clear
separation of scales does take place and only in this case are the present
techniques really accurate. The appeareance of pseudo Goldstone bosons
(abundant in models of dynamical breaking) may thus jeopardize our
conclusions,
as they bring a relatively light scale into the game (typically even
lighter than the Fermi scale). In fact, for the observables we
consider their contribution is not too important, unless they are
extremely light. For instance a pseudo-Goldstone boson of 100 GeV
can be accommodated without much trouble, as we shall later see.

The four-fermion operators we have just alluded to can involve either four
ordinary
quarks or leptons (but we will see that dimensional counting suggests that
their contribution will be irrelevant at present energies with the
exception of those containing the top quark), or two new (heavy)
fermions and two
ordinary ones. This scenario is quite natural in several extended
technicolor
(ETC) or top condensate (TopC) models\cite{ETC,TopC}, in which the underlying
dynamics is characterized
by a scale $M$. At scales $\mu < M$ the dynamics can be modelled by
four-fermion operators (of either technifermions in ETC models,
or ordinary fermions of the third family in TopC models). We perform a
classification\footnote{In the case of ordinary fermions and leptons,
four-fermion operators have been studied in \cite{4FC}. To our knowledge
a complete analysis when additional fields beyond those
present in the Standard Model are present has not been presented in
the literature before.}
of these operators. We shall concentrate
in the case where technifermions
appear in ordinary representations of $SU(2)_L\times SU(3)_c$ (hypercharge
can be arbitrary).  The classification will then be exhaustive. We shall
discuss other representations as well, although we shall consider
custodially preserving operators only, and only those operators which are
relevant for our purposes.

As a matter of principle we have tried not to make
any assumptions regarding the actual way different generations
are embedded in the extended interactions. In practice, when presenting
our numerical plots and figures, we are assuming that the
appropriate group-theoretical factors are similar for
all three generations of physical fermions.

It has been our purpose in this paper to be as general as possible, not
advocating or trying to put forward any particular theory.
Thus, the analysis
may, hopefully, remain useful beyond the models we have just used to
motivate the problem. We hope to convey to the reader our belief
that a systematic approach based on four-fermion operators
and the effective lagrangian treatment can be very useful.

\section{The matter sector}

\label{S-17.6a}

Appelquist, Bowick, Cohler and Hauser established some time ago a list of $%
d=4$ operators\cite{ABCH}. These are
the operators of lowest dimensionality which are non-universal. In other
words, their coefficients will contain information on whatever mechanism
Nature has chosen to make quarks and leptons massive. Of course operators
of dimensionality 5, 6 and so on will be generated at the same time. We
shall turn to these later. We have reanalysed all possible independent
operators of $d=4$ (see the discussion in appendix A) and we find the
following ones
\begin{eqnarray}
\Ap[1,4]&=&\ri \bar q_L\U (\Ds\U)^\da q_L  \label{2.6b} \\
\Ap[2,4]&=&\ri\bar q_R \Ud (\Ds \U)q_R \\
\Ap[3,4]&=&\ri \bar q_L(\Ds\U)\tau^3 \Ud q_L-\ri \bar q_L \U\tau^3(\Ds\U%
)^\da q_L \\
\Ap[4,4]&=&\ri \bar q_L \U\tau^3 \Ud (\Ds\U)\tau^3 \Ud q_L \\
\Ap[5,4]&=&\ri \bar q_R\tau^3\Ud (\Ds\U) q_R-\ri\bar q_R(\Ds\U)^\da \U %
\tau^3 q_R \\
\Ap[6,4]&=&\ri\bar q_R \tau^3 \Ud(\Ds\U) \tau^3 q_R \\
\Ap[7,4]&=&\ri\bar q_L\U\tau^3 \Ud\Ds q_L-\ri \bar q_L \Ds^\da \U\tau^3\Ud %
q_L  \label{2.6c} \\
\Ap[,4]{}^{\prime}&=&\ri\bar q_R \tau^3 \Ds q_R-\ri \bar q_{R}\Ds%
^\dagger\tau^3 q_{R} ,  \label{2.6d}
\end{eqnarray}
where it is understood that $(\Ds\U)^\da\equiv\gamma_\mu(D^\mu\U)^\da$.
Each operator is accompanied by a coefficient $\delta^\prime, \delta_1,
\delta_2, \ldots \delta_7$, thus, up to ${\cal O}(p^4)$, our
effective lagrangian is\footnote{%
Although there is only one derivative in  (\ref{30.6b}) and thus this is
a misname,
we stick to the same notation here as in the purely bosonic
effective lagrangian}
\begin{equation}
{\cal L}_{{\rm eff}}= \delta^\prime \Ap[,4]{}^{\prime}+ \sum_{i=1}^7
\delta_i \Ap[i,4].  \label{30.6b}
\end{equation}
In the above, $D_{\mu}U$ is defined in~(\ref{30.6a}) whereas
\begin{eqnarray}
D_{\mu} q_{L}&=&\left(\p_{\mu}+\ri g{\frac{\vec\tau}{2}} \cdot \vec W_{\mu} +%
{\ri g^{\prime} Y} B_{\mu} \right) q_{L}, \\
D_{\mu} q_{R}&=&\left(\p_{\mu}+\ri g^{\prime}{\frac{\tau_{3}}{2}} B_{\mu} +{%
\ri g^{\prime} Y} B_{\mu} \right) q_{R} .
\end{eqnarray}
where $Y=\mathbf{I}/6$ for quarks and $Y=-\mathbf{I}/2$ for leptons.
This list differs from the one in \cite{ABCH} by the presence of the last
operator (\ref{2.6d}). It will turn out, however, that $\delta^\prime$ does
not contribute to any observable.
All these operators are invariant under local $SU(2)_L\times U(1)_Y$
transformations.

This list includes both custodially preserving operators, such as $\Ap[1,4]$
and $\Ap[2,4]$, and custodially breaking ones, such as $\Ap[,4]{}^{\prime}$
and $\Ap[3,4]$ to $\Ap[7,4]$. In the purely bosonic part of the
effective lagrangian (\ref{effobl}), the first (universal) operator and the
one accompanying $a_1$ are custodially preserving, while those going with $%
a_0$ and $a_8$ are custodially breaking. E.g., $a_0$ parametrizes the
contribution of the new physics to the $\Delta \rho$ parameter. If the
underlying physics is custodially preserving only $\delta_1$ and $\delta_2$
will get non-vanishing contributions\footnote{Of course hypercharge breaks
custodial symmetry, since only a subgroup of $SU(2)_R$ is gauged. Therefore,
{\it all} operators involving right-handed fields
break custodial symmetry. However, there is still
a distinction between those operators whose structure is formally custodially
invariant (and custodial symmetry is broken only through the coupling to the
external gauge field) and those which would not be custodially
preserving even if the full  $SU(2)_R$ were gauged.}.

The operator $\Ap[7,4]$ deserves some comments. By using the equations of
motion it can be reduced to the mass term (\ref{mass})
\begin{equation}
\delta_7\, v\bar q_L U (y\tau_3+y_3)q_R + {\rm h. c.} .
\end{equation}
However this procedure is, generally speaking, only justified if the matter
fields appear only as external legs. For the time being we shall keep $%
\Ap[7,4]$ as an independent operator and in the next section we shall
determine its value in the minimal Standard Model after integrating out a
heavy Higgs. We shall see that, after imposing that physical on-shell
fields have unit residue, $\delta_7$ does drop from all physical predictions.

What is the expected size of the $\delta_i$ coefficients in the minimal
Standard Model? This question is easily answered if we take a look at the
diagrams that have to be computed to integrate out the Higgs field
(figure \ref{fig-3}). Notice that the calculation
is carried out in the non-linear variables $U$%
, hence the appearance of the unfamiliar diagram e). Diagram d) is actually
of order $1/M_H^2$, which guarantees the gauge independence of the effective
lagrangian coefficients. The diagrams are obviously proportional to $y^2$,
$y$ being
a Yukawa coupling, and also to $1/16\pi^2$, since they originate
from a one-loop calculation. Finally, the screening theorem shows that they
may depend on the Higgs mass only logarithmically, therefore
\begin{equation}
\delta_i^{{\rm SM}}\sim {\frac{y^2}{{16\pi^2}}} \log{\frac{M_H^2}{M_Z^2}}.
\end{equation}
These dimensional considerations show that the vertex corrections are
only sizeable for third
generation quarks.

In models of
dynamical symmetry breaking, such as TC or ETC,
we shall have new contributions to the $\delta_i$ from
the new physics (which we shall later parametrize with four-fermion
operators). We have several new scales at our disposal. One is $M$, the
mass normalizing dimension six four-fermion operators. The other can be
either $m_b$ (negligible, since $M$ is large), $m_t$, or the dynamically
generated mass of the techniquarks $m_Q$ (typically of order $\Lambda_{TC}$,
the scale associated to the interactions triggering the breaking of the
electroweak group). Thus we can get a contribution of order
\begin{equation}
\delta_i^{\rm Q}\sim \frac{1}{16\pi^2} \frac{m_Q^2}{M^2}\log\frac{m_Q^2}{M^2}.
\end{equation}
While $m_Q$ is, at least naively,  expected to be $\simeq \Lambda_{TC}$
and therefore  similar  for all flavours, there should be a hierarchy
for
$M$. As will be discussed in the following sections, the scale $M$ which is
relevant for the mass generation (encoded in the only dimension 3 operator
in the effective lagrangian), via techniquark condensation and ETC
interaction exchange (figure~\ref{fig-4}),
is the one normalizing
chirality flipping operators. On the contrary, the scale normalizing
dimension 4 operators in the
effective theory is the one that
normalizes chirality preserving operators. Both scales need not be exactly
the same, and one may envisage a situation with relatively light
scalars present where the former can be much lower. However, it is
natural to expect that
$M$ should at any rate
be smallest for the third generation.
Consequently the contribution to the $\delta_i$'s from
the third generation should be largest.
\fig{}{0cm}{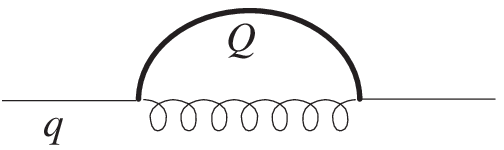}{Mechanism generating quark masses through
the exchange of a ETC particle.}{fig-4}

We should also discuss dimension 5, 6, etc operators and why we need not
include them in our analysis. Let us write some operators of dimension 5:
\bea
&&\bar q_L\hW\U q_R +\mbox{h. c.} ,\label{op1'}\\
&&\bar q_L\U \hB q_R +\mbox{h. c.} ,\label{op5'}\\
&&\bar q_L\si^{\mu\nu}D^\da{}_{[\mu} D_{\nu]}\U q_R-
              \bar q_L\si^{\mu\nu}
D_{[\nu}\U D_{\mu]} q_R+\mbox{h. c.} ,\label{op7'}\\
&&\bar q_L \U D^2 q_R+\mbox{h. c.} ,\label{op11'}\\
&&\bar q_L D^\da_\mu \U\tau^3\Ud (D^\mu \U)q_R-
\bar q_L \U\tau^3\Ud (D^\mu \U) D_\mu q_R+\mbox{h. c.};\label{op18'}\\
&&\ldots \nn
\eea
where we use the notation $\hW\equiv\ri g \si^{\mu\nu} W_{\mu\nu}$,
$\hB\equiv\ri g' \si^{\mu\nu} B_{\mu\nu}$.
These are a few of a
long list of about 25 operators, and this including only the ones
contributing to the $ffZ$ vertex. All these operators are however chirality
flipping and thus their contribution to the amplitude must be suppressed by
one additional power of the fermion masses. This makes their study
unnecessary
at the present level of precision. Similar considerations apply to operators
of dimensionality 6 or higher.

\section{The effective theory of the Standard Model}
\label{31.7-1}

In this section we shall obtain the values of the coefficients $\delta _i$
in the minimal Standard Model. The appropriate effective coefficients for
the oblique corrections $a_i$ have been obtained previously by several
authors\cite{EH,EM,all}. Their values are
\begin{equation}
a_0=\frac 1{16\pi ^2}\frac 38\left( \frac 1{\hat{\epsilon}}-\log \frac{M_H^2%
}{\mu ^2}+\frac 56\right)
\end{equation}
\begin{equation}
a_1=\frac 1{16\pi ^2}\frac 1{12}\left( \frac 1{\hat{\epsilon}}-\log \frac{%
M_H^2}{\mu ^2}+\frac 56\right)
\end{equation}
\begin{equation}
a_8=0.
\end{equation}
where $1/\hat{\ep }\equiv 1/\ep-\ga_E+\log 4\pi $. We use dimensional
regularization with a spacetime dimension $4-2\ep$.

We begin by writing the Standard Model in terms of the non-linear variables $%
U$. The matrix
\begin{equation}
{\cal M}=\sqrt{2}(\tilde\Phi, \Phi),
\end{equation}
constructed with the Higgs doublet, $\Phi$ and its conjugate, $%
\tilde\Phi\equiv \ri \tau^2 \Phi^{*}$, is rewritten in the form
\begin{equation}
{\cal M}=(v+\rho)U, \qquad U^{-1}=U^\dagger ,
\end{equation}
where $\rho$ describe the `radial' excitations around the v.e.v. $v$.
Integrating out the field $\rho$ produces an effective lagrangian of the
form (\ref{effobl}) with the values of the $a_i$ given above (as well as
some other pieces not shown there). This functional integration also
generates the vertex corrections~(\ref{30.6b}).

We shall determine the $\delta_i$ by demanding that the renormalised
one-particle irreducible Green functions (1PI), $\hat\Ga$, are the same (up
to some power in the external momenta and mass expansion) in both, the
minimal Standard Model and the effective lagrangian. In other words, we
require that \beq
\De \hat\Ga=0,
\label{match-2}
\eeq
where throughout this section
\beq
\De\Ga\equiv \Ga_{{\rm SM}}-\Ga_{{\rm eff}},
\eeq
and the hat denotes renormalised quantities. This procedure is known as
matching. It goes without saying that in doing so the same renormalization
scheme must be used. The on-shell scheme is particularly well suited to
perform the matching and will be used throughout this paper.

One only needs to worry about SM diagrams that are not present in the
effective theory; namely, those containing the Higgs. The rest of the
diagrams give exactly the same result, thus dropping from the matching. In
contrast, the diagrams containing a Higgs propagator are described by local
terms (such as $\Ap[1,4]$ through $\Ap[7,4]$) in the effective theory, they
involve the coefficients $\delta_i$, and give rise to the Feynman rules
collected in appendix~\ref{app-3.6b}.

Let us first consider the fermion self-energies. There is only one 1PI
diagram with a Higgs propagator (see figure~\ref{fig-3}).
\fig{ }{0cm}{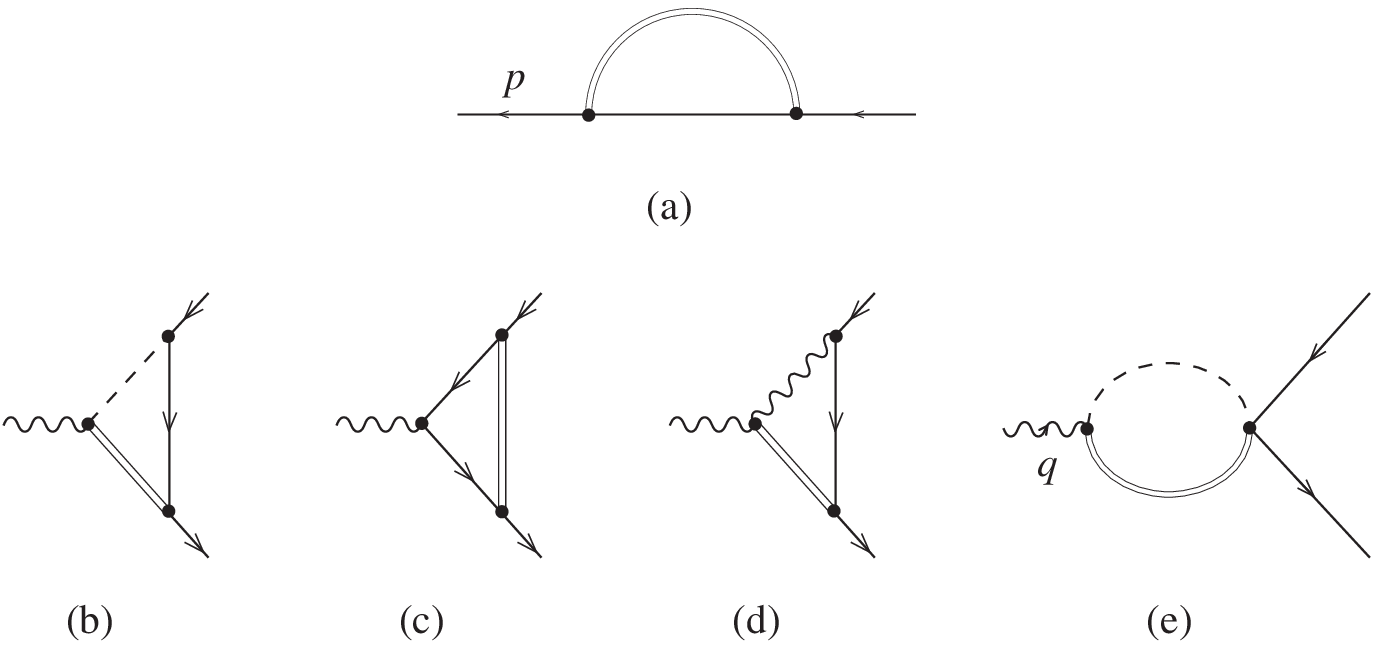}{The
diagrams relevant for the matching of the fermion self-energies and vertices
(counterterm diagrams are not included). Double lines
represent the Higgs, dashed lines
the Goldstone bosons, and wiggly lines
the
gauge bosons.
}{fig-3} A straightforward calculation gives
\beq
\Si^f_{{\rm SM}} =-{\frac{y_f^2}{16\pi^2}}\left\{ \ps\left[ {\frac{1}{2}}\eph
-{\frac{1}{2}}\log{\frac{M_H^2}{\mu^2}}+{\frac{1}{4}} \right] +m_f\left[ \eph
-\log{\frac{M_H^2}{\mu^2}}+1 \right] \right\}. \label{SM5}
\eeq
$\De\Si^f$ can be computed by subtracting~(\ref{new-1}), (\ref{new-2})
from~(\ref{SM5}).

Next, we have to renormalise the fermion self-energies. We introduce the
following notation \beq
\De Z\equiv Z_{{\rm SM}}-Z_{{\rm eff}}= \de Z_{{\rm SM}}-\de Z_{{\rm eff}},
\eeq
where $Z_{{\rm SM}}$ ($Z_{{\rm eff}}$) stands for any renormalization
constant of the SM (effective theory). To compute $\De\hat\Si^f$, we simply
add to $\De\Si^f$ the counterterm diagram~(\ref{17.6a}) with the
replacements $\de Z^f_{V,A} \to \De
Z^f_{V,A}$ and $\de m_{f}\to \De m_{f}$. This, of course, amounts to
eqs.~(\ref{17.6d}), (\ref{17.6e}) and~(\ref{17.6f}) with the same
replacements. From~(\ref{9.6a}), (\ref{17.6b}) and (\ref{17.6c}) (which also
hold for $\De Z$, $\De m$ and $\De \Si$) one can express $\De Z^f_{V,A}$ and
$\De m_{f}/m_{f}$ in terms of the bare fermion self-energies and finally
obtain $\De\hat\Si^f$. The result is \bea
\De\hat\Si^d_{A,V,S}&=&0\\\De\hat\Si^u_{A}&=&0\\\De\hat\Si^u_{V,S}&=&4
\de_7-{
\frac{ 1}{16\pi^2}}{\frac{y_u^2-y_d^2}{2}} \left[\eph-\log{\frac{M_H^2}{\mu^2
}}+{\frac{1}{2}} \right]. \label{17.6g} \eea
We see from~(\ref{17.6g}) that the matching conditions, $\De
\hat\Si^u_{V,S}=0$, imply \beq
\de_7={\frac{ 1}{16\pi^2}}{\frac{y_u^2-y_d^2}{8}} \left[\eph-\log{\frac{M_H^2
}{\mu^2}}+{\frac{1}{2}} \right]. \label{SM-de7} \eeq
The other matchings are satisfied automatically and do not give any
information.

Let us consider the vertex $ffZ$. The relevant diagrams are shown in
figure~\ref{fig-3} (diagrams b--e).  We shall only collect the
contributions proportional to $\ga_{\mu}$ and $\ga_{\mu}\ga_{5}$. The result
is \beq
\Ga_\mu^{ffZ}= -{\frac{\ri}{16\pi^2}}{\frac{y_f^2}{2}}\ga_\mu\left\{ v_f
\left(\eph-\log{\frac{M_H^2}{\mu^2}}+{\frac{1}{2}}\right)- 3a_f\, \ga_5
\left(\eph-\log{\frac{M_H^2}{\mu^2}}+{\frac{11}{6}}\right) \right\}. \eeq
By subtracting the diagrams~(\ref{newN1d1}) and~(\ref{newN1d2}) from $\Ga
_\mu^{ffZ}$ one gets $\De\Ga_\mu^{ffZ}$. Renormalization requires that we
add the counterterm diagram~(\ref{17.6h}) where, again, $\de Z\to \De Z$.
One can check that both $\De\ZZp-\De\ZZ$ and $\De\ZZgap-\De\ZZga$ are
proportional to $\De\Si_{Z\ga}(0)$, which turns out to be zero. Hence the
only relevant renormalization constants are $\De Z_V^f$ and $\De Z_A^f$.
These renormalization constant have already been determined. One obtains for
$\De\hat\Ga^{ffZ}_\mu$ the result
\bea
\De\hat\Ga^{ddZ}_\mu&=&- {\frac{\ri e}{2 s_Wc_W}}\ga_\mu\left\{
\left[\Half (\de_{1}-\de_{4}-\de_{2}-\de_{6})+\de_{3}+\de_{5}\right] \phantom{
\left[ \eph-\log{\frac{M_H^2}{\mu^2}}+{\frac{5}{2}} \right]} \right.
\nn\\
&-&\left. \ga_5\left[ {\frac{1}{16\pi^2}}{\frac{y_d^2}{2}}\left( \eph-\log{
\frac{M_H^2}{\mu^2}}+{\frac{5}{2}} \right)+ \Half (\de_1 -  \de_4 +
\de_{2}
+\de_6)+\de_{3}-\de_5 \right]\right\} \\
\De\hat\Ga^{uuZ}_\mu&=&- {
\frac{\ri e}{2 s_Wc_W}}\ga_\mu
\left\{ \left[\Half (\de_{1}-\de_{4}-\de_{2}-
\de_{6})-\de_{3}-\de_{5}\right] \phantom{\left[ \eph-\log{\frac{M_H^2
}{\mu^2}}+{\frac{5}{2}} \right]} \right. \nn\\&-&\left. \ga_5\left[ -{\frac{1
}{16\pi^2}}{\frac{y_u^2}{2}}\left( \eph-\log{\frac{M_H^2}{\mu^2}}+{\frac{5}{2
}} \right)- \Half (\de_{1}-\de_{4}+\de_{2}+\de_{6})+\de_{3}-\de_{5}
\right]\right\}, \eea
where use has been made of eq.~(\ref{SM-de7}). The matching condition, $\De\hat\Ga^{ffZ}_\mu
= 0$ implies \bea
\de_1-\de_4&=&-{\frac{1}{16\pi^2}}{\frac{y_u^2+y_d^2}{4}}\left( \eph-\log{
\frac{M_H^2}{\mu^2}}+{\frac{5}{2}} \right)\label{SMmt1}\\\de_2+\de_6&=&-{
\frac{1}{16\pi^2}}{\frac{y_u^2+y_d^2}{4}}\left( \eph-\log{\frac{M_H^2}{\mu^2}
}+{\frac{5}{2}} \right)\label{SMmt2}\\\de_3&=&\phantom{-}{\frac{1}{16\pi^2}}{
\frac{y_u^2-y_d^2}{4}}\left( \eph-\log{\frac{M_H^2}{\mu^2}}+{\frac{5}{2}}
\right)\label{SMmt3}\\\de_5&=&-{\frac{1}{16\pi^2}}{\frac{y_u^2-y_d^2}{4}}
\left( \eph-\log{\frac{M_H^2}{\mu^2}}+{\frac{5}{2}} \right) .\label{SMmt4}
\eea

To determine completely the $\de_i$ coefficients we need to consider the
vertex $udW$. The relevant diagrams are analogous
to those of figure~\ref{fig-3}.
A straightforward calculation gives
\bea
\De\hat\Ga_\mu^{udW}&=&{\frac{\ri e}{4\sqrt{2} s_W}}\; \ga_\mu\left\{\left[ {
\frac{y_u y_d}{16\pi^2}}\left(\eph-\log{\frac{M_H^2}{\mu^2}}+{\frac{5}{2}}
\right) +2\de_2-2\de_6 \right](1+\ga_5)\right.\nn\\&&\left.-\left[{\frac{
y_u^2+y_d^2}{16\pi^2}}\Half \left( \eph-\log{\frac{M_H^2}{\mu^2}}+{\frac{5}{2
}} \right)+2\de_1+2\de_4\right](1-\ga_5)\right\}. \eea
The matching condition $\De\hat\Ga_\mu^{udW}=0$ amounts to the following set
of equations \bea
\de_2-\de_6&=&-{\frac{1}{16\pi^2}}{\frac{y_u y_d}{2}} \left( \eph-\log{\frac{
M_H^2}{\mu^2}}+{\frac{5}{2}} \right)\\\de_1+\de_4&=&-{\frac{1}{16\pi^2}}{
\frac{y_u^2+ y_d^2}{4}} \left( \eph-\log{\frac{M_H^2}{\mu^2}}+{\frac{5}{2}}
\right) \eea
Combining these equations with eqs.~(\ref{SMmt1}, \ref{SMmt2}) we finally
get \bea
\de_1&=&-{\frac{1}{16\pi^2}}{\frac{y_u^2+ y_d^2}{4}} \left( \eph-\log{\frac{
M_H^2}{\mu^2}}+{\frac{5}{2}} \right)\\\de_2&=& -{\frac{1}{16\pi^2}}{\frac{
(y_u+y_d)^2}{8}} \left( \eph-\log{\frac{M_H^2}{\mu^2}}+{\frac{5}{2}} \right)
\\\de_4&=&0\\\de_6&=&-{\frac{1}{16\pi^2}}{\frac{(y_u-y_d)^2}{8}} \left( \eph
-\log{\frac{M_H^2}{\mu^2}}+{\frac{5}{2}} \right) \eea
This, along with eqs.~(\ref{SMmt3}, \ref{SMmt4}) and eq.~(\ref{SM-de7}),
is our final answer. These results coincide, where the comparison is possible,
with those obtained in \cite{DG} by functional methods. It is interesting to
note that it has not been necessary to consider the matching of the vertex $
ff\ga$.

We shall show explicitly that $\de_7$ drops from the $S$
matrix element corresponding to $Z\to f \bar f$.
It is well known that the renormalised $u$-fermion self-energy has residue
$1+\de_\res$, where $\de_\res$ in given in eq.~(\ref{17.6j}) of appendix D.
Therefore, in order to evaluate $S$-matrix elements involving external $u$
lines at one-loop, one has to multiply the corresponding amputated Green
functions by a factor $1+n\,\de_\res /2$, where $n$ is the number on external
$u$-lines (in the case under consideration $n=2$). One can check that when
this factor is taken into account, the $\de_{7}$ appearing in the
renormalised S-matrix vertex are cancelled.

We notice that $\delta_1$ and $\delta_2$ indeed correspond to custodially
preserving operators, while $\delta_3$ to $\delta_6$ do not. All these
coefficients (just as $a_0$, $a_1$ and $a_8$) are ultraviolet divergent.
This is so because the Higgs particle is an essential ingredient to
guarantee the renormalizability of the Standard Model. Once this is removed,
the usual renormalization process (e.g. the on-shell scheme) is not enough
to render all ``renormalised" Green functions finite. This is why the bare
coefficients of the effective lagrangian (which contribute to the
renormalised Green functions either directly or via counterterms) have to be
proportional to $1/\epsilon$ to cancel the new divergences. The coefficients
of the effective lagrangian are manifestly gauge invariant.

What is the value of these coefficients in other theories with
elementary scalars and Higgs-like mechanism? This issue has
been discussed in some detail in \cite{ciafaloni} in the context
of the two-Higgs doublet model, but it can actually be
extended to supersymmetric theories (provided of course scalars
other than the CP-even Higgs can be made heavy enough, see e.g. \cite{mjh}).
It
was argued there that non-decoupling effects are exactly the same as in the
minimal Standard Model, including the constant non-logarithmic piece.
Since the
$\delta_i$ coefficients contain all the non-decoupling effects
associated to the Higgs particle at the first non-trivial order in the
momentum or mass expansion, the low energy effective theory will
be exactly the same.

\section{Observables}

The decay width of $Z\rightarrow f\bar{f}$ is described by
\begin{equation}
\Gamma _f\equiv \Gamma \left( Z\rightarrow f\bar{f}\right) =4n_c\Gamma
_0\left[ \left( g_V^f\right) ^2R_V^f+\left( g_A^f\right) ^2R_A^f\right] ,
\end{equation}
where $g_V^f$ and $g_A^f$ are the effective electroweak couplings as defined
in \cite{yellow} and $n_c$ is the number of colours of fermion $f$. The
radiation factors $R_V^f$ and $R_A^f$ describe the final state QED and QCD
interactions \cite{yellow2}. For a charged lepton we have
\begin{eqnarray*}
R_V^l &=&1+\frac{3\bar{\alpha}}{4\pi }+{\cal O}\left( \bar{\alpha}^2,\left(
\frac{m_l}{M_Z}\right) ^4\right) , \\
R_A^l &=&1+\frac{3\bar{\alpha}}{4\pi }-6\left( \frac{m_l}{M_Z}\right) ^2+
{\cal O}\left( \bar{\alpha}^2,\left( \frac{m_l}{M_Z}\right) ^4\right) ,
\end{eqnarray*}
where $\bar{\alpha}$ is the electromagnetic coupling constant at the scale $%
M_Z$ and $m_l$ is the final state lepton mass

The tree-level width $\Gamma _0$ is given by
\begin{equation}
\Gamma _0=\frac{G_\mu M_Z^3}{24\sqrt{2}\pi }.
\end{equation}
If we define
\begin{equation}
\rho _f\equiv 4\left( g_A^f\right) ^2,
\end{equation}
\begin{equation}
\bar{s}_W^2\equiv \frac{I_f^3}{2Q_f}\left( 1-\frac{g_V^f}{g_A^f}\right) ,
\end{equation}
we can write
\begin{equation}
\Gamma _f=n_c\Gamma _0\rho _f\left[ 4\left( I_f^3-2Q_f\bar{s}_W^2\right)
^2R_V^f+R_A^f\right] .
\end{equation}
Other quantities which are often used are $\Delta \rho _f$, defined through
\begin{equation}
\rho _f\equiv \frac 1{1-\Delta \rho _f},
\end{equation}
the forward-backward asymmetry $A_{FB}^f$
\begin{equation}
A_{FB}^f=\frac 34A^eA^f,
\end{equation}
and $R_b$
\begin{equation}
R_b=\frac{\Gamma _b}{\Gamma _h},
\end{equation}
where
\[
A^f\equiv \frac{2g_V^fg_A^f}{\left( g_A^f\right) ^2+\left( g_V^f\right) ^2},
\]
and $\Gamma _b$, $\Gamma _h$ are the b-partial width  and
total hadronic width, respectively (each of them, in turn, can be expressed
in terms of the appropriate effective couplings). As we see, nearly all of
$Z$ physics can
be described in terms of $g_A^f$ and $g_V^f$. The box contributions to the
process $e^{+}e^{-}\rightarrow f\bar{f}$ are not included in the analysis
because they are negligible and they cannot be incorporated as
contributions to effective electroweak neutral current couplings anyway.

We shall generically denote these effective couplings by $g^f$. If we
express the value they take in the Standard Model by $g^{f\left( SM\right) }$%
, we can write a perturbative expansion for them in the following way
\begin{equation}
g^{f\left( {\rm SM}\right) }=
g^{f\left( 0\right) }+g^{f\left( 2\right) }+\bar{g}%
^f( {a}^{\rm SM}) +\hat{g}^f( {\delta }^{\rm SM}),
\label{gsm}
\end{equation}
where $g^{f\left( 0\right) }$ are the tree-level expressions for these form
factors, $g^{f\left( 2\right) }$ are the one-loop contributions which do not
contain any Higgs particle as internal line in the Feynman graphs. In the
effective lagrangian language they are generated by the quantum corrections
computed by operators such as (\ref{mass}) or the first operator on the
r.h.s. of (\ref{effobl}). On the other hand, the Feynman diagrams containing
the Higgs particle contribute to $g^{f\left( SM\right) }$ in a twofold way.
One is via the ${\cal O}(p^2)$ and ${\cal O}(p^4)$ Longhitano effective
operators (\ref{effobl}) which depend on the $a_i$ coefficients, which are
Higgs-mass dependent, and thus give a Higgs-dependent oblique correction to
$g^{f\left( SM\right) }$, which is denoted by $\bar{g}^f$. The other one is
via genuine vertex corrections which depend on the $\delta_i$. This
contribution is denoted by $\hat{g}^f$.

The tree-level value for the form factors are
\begin{equation}
g_V^{f\left( 0\right) }=I_f^3-2s_W^2Q_f,\qquad g_A^{f\left(
0\right) }=I_f^3.
\end{equation}

In a theory X, different from the minimal Standard Model, the effective
form factors will take values $g^{f\left({\rm X}\right) }$, where
\begin{equation}
g^{f\left( {\rm X}\right) }=
g^{f\left( 0\right) }+g^{f\left( 2\right) }+\bar{g}%
^f( {a}^{\rm X}) +\hat{g}^f( {\delta }^{\rm X}),
\label{g4q}
\end{equation}
and the ${a}^{\rm X}$ and ${\delta }^{\rm X}$ are effective coefficients
corresponding to theory X.

Within one-loop accuracy in the symmetry breaking sector (but with arbitrary
precision elsewhere), $\bar{g}^f$ and $\hat{g}^f$ are linear functions
of their arguments and thus we have
\begin{equation}
g^{f\left( {\rm X}\right) }=g^{f\left({\rm  SM}\right) }+\bar{g}^f ( {a}^{\rm X}-
{a}^{\rm SM}) +\hat{g}^f( {\delta }^{\rm X}-{\delta }^{\rm SM}) .
\end{equation}

The expression for $\bar{g}^{f }$ in terms of $a_i$ was already given in (%
\ref{gagvobl}) and (\ref{gagvvv}). On the other hand
from appendix B we learn that
\begin{equation}
\hat{g}_V^f\left( \delta _1,\cdots ,\delta _6\right) =I_f^{3} \left( \delta
_1-\delta _4-\delta _2-\delta _6\right) -\delta _3-\delta _5,
\end{equation}
\begin{equation}
\hat{g}_A^f\left( \delta _1,\cdots ,\delta _6\right) =I_f^{3} \left( \delta
_1-\delta _4+\delta _2+\delta _6\right) -\delta _3+\delta _5,  \label{1}
\end{equation}

In the minimal Standard Model all the Higgs dependence at the one loop level
(which is the level of accuracy assumed here) is logarithmic and
is contained in the $a_i$ and $\delta_i$ coefficients.
Therefore one
can easily construct linear combinations of observables where the leading
Higgs dependence cancels. These combinations allow for
a test of the minimal Standard Model independent of the actual
value of the Higgs mass.

Let us now review the comparison with current electroweak data
for theories with dynamical symmetry breaking. Some confusion seem to
exist on this point so let us try to analyze this issue critically.

A first difficulty arises from the fact that at the $M_Z$ scale
perturbation theory is not valid in theories with dynamical breaking
and the contribution
from the symmetry breaking sector must be estimated in the framework
of the effective theory, which is non-linear and non-renormalizable.
Observables will depend on some subtraction scale.
(Estimates based on dispersion relations
and resonance saturation amount, in practice, to the same, provided
that due attention is paid to the scale dependence introduced by the
subtraction in the dispersion relation.)

A somewhat related problem is that, when making use of the
variables $S,T$ and $U$\cite{PT}, or $\epsilon_1 , \epsilon_2$
and $\epsilon_3$\cite{epsilon}, one often sees in the
literature bounds on possible ``new physics"
in the symmetry breaking sector without actually removing the
contribution from the Standard Model higgs that
the ``new physics" is supposed to replace (this is not the case e.g. in
\cite{PT} where this issue is discussed with some care). Unless the contribution
from the ``new physics"  is enormous, this is a flagrant
case of double counting, but it is easy to understand why this
mistake is made: removing the Higgs makes the Standard Model
non-renormalizable and the observables of the Standard Model
without the Higgs depend on some arbitrary subtraction scale.

In fact the two sources of arbitrary subtraction scales
(the one originating from the removal of the Higgs and
the one from the effective action treatment)
are one an the same and the problem can be dealt with
the help of the coefficients of higher dimensional
operators in the effective
theory (i.e. the $a_i$ and $\delta_i$).
 The dependence
on the unknown subtraction scale is absorbed in the coefficients
of higher dimensional operators and traded by the scale of the
``new physics". Combinations of observables
can be built where this scale (and the associated renormalization
ambiguities) drops. These combinations
allow for a test of the ``new physics" independently of the
actual value of its characteristic scale. In fact they are
the same combinations of observables where the
Higgs dependence drops in the minimal Standard Model.

A third difficulty in making a fair comparison of models
of dynamical symmetry breaking with experiment lies in the vertex
corrections.
If we analyze the
lepton effective couplings
$g_A^l$ and $g_V^l$, the minimal Standard
Model predicts very small vertex corrections arising from the symmetry
breaking sector anyway and it is consistent to ignore them and concentrate
in the oblique corrections. However, this is not the situation
in dynamical symmetry breaking models. We will see in the next sections
that for the second and third generation vertex corrections
can be sizeable. Thus if we want to compare experiment to
oblique corrections in models of dynamical breaking
we have to concentrate on electron couplings only.

\fig{}{0cm}{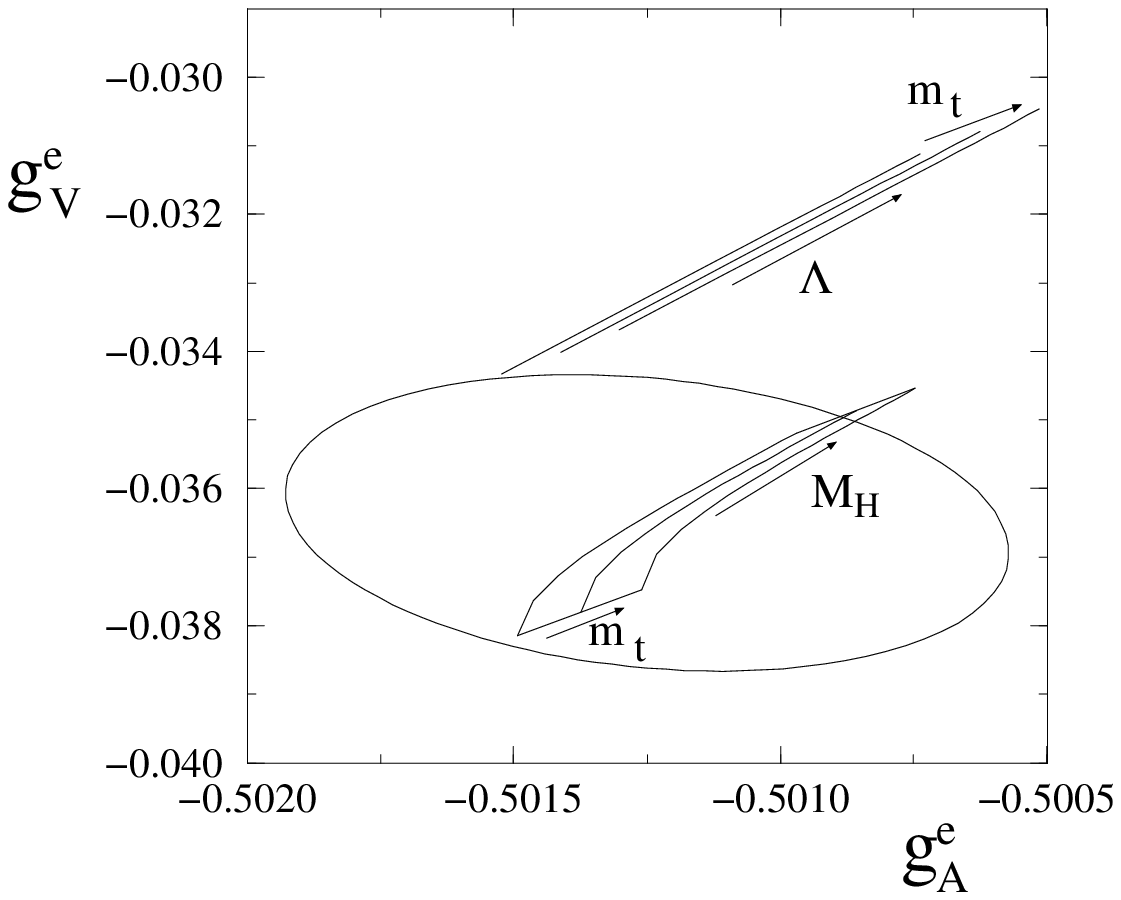}{The $1-\sigma$ experimental region in the
$g_A^e-g_V^e$ plane. The Standard Model predictions as a function
of $m_t$ ($170.6 \le m_t \le 180.6$ GeV) and $M_H$ ($70\le M_H\le 1000$ GeV)
are shown (the middle line corresponds to the central value $m_t=175.6 $
GeV). The predictions of a QCD-lke
technicolor theory with $n_{TC}n_D=8$ and
degenerate technifermion masses are shown as straight lines (only oblique
corrections are included). One moves along the straight lines by changing
the scale $\Lambda$. The three lines correspond to the extreme and central
values for $m_t$. Recall that the precise location anywhere on the straight
lines (which definitely do intersect the $1 - \sigma$ region) depends
on the renormalization procedure and thus is not predictable within the
non-renormalizable effective theory. In
addition the technicolor prediction should be considered accurate only
at the 15\% level due to the theoretical uncertainties
discussed in the text (this  error is at any rate smaller than the one
associated to the uncertainty in $\Lambda$). Notice that
the oblique corrections, in the case of degenerate masses, are independent
of the value of the technifermion mass. Assuming universality
of the vertex corrections reduces the error bars by about a factor one-half
and leaves technicolor predictions outside the $1 - \sigma$ region.}{fig-1}

In figure~\ref{fig-1} we see the prediction of
the minimal Standard Model for $170.6
<m_t< 180.6$ GeV and $70 <M_H< 1000$ GeV including
the leading two-loop corrections%
\cite{yellow2}, falling nicely within the experimental
$1 - \sigma$ region for the electron effective couplings.
In this and in subsequent plots we present the data form the combined
four LEP experiments only. What is the actual prediction for a theory with
dynamical symmetry breaking?
The straight solid lines
correspond to the prediction of a QCD-like technicolor
model with $n_{TC}=2$ and $n_D=4$ (a one-generation model)
in the case where all
technifermion masses are assumed to be equal (we follow \cite{DEH},
see \cite{other} for
related work) allowing the same variation for the top
mass as in the Standard Model. We do not take
into account here the contribution of potentially present
pseudo Goldstone bosons, assuming that they can be made heavy
enough.
The corresponding values for the $a_i$ coefficients in
such a model are given in appendix E and are derived
using chiral quark model techniques and chiral perturbation theory.
They are scale dependent in such a way as to make
observables finite and unambiguous, but of course observables
depend in general on the scale of ``new physics" $\Lambda$.

We move along the straight lines by
changing the scale $\Lambda$.
It would appear at first sight that one needs to go
to unacceptably low values
of the new scale to actually penetrate the $1-\sigma$ region, something
which looks unpleasant at
first sight (we have plotted the part of the line for $100 \le
\Lambda \le 1500$ GeV), as one expects $\Lambda\sim \Lambda_\chi$. In fact
this is not necessarily so. There is no real prediction of the
effective theory {\em along} the straight lines, because
only combinations which are $\Lambda$-independent are predictable.
As for the location not {\em along} the line,
but {\em of} the line itself it is in principle calculable in the effective
theory,
but of course subject to the uncertainties of the model one relies upon,
since we are dealing with a strongly coupled theory.
(We shall use chiral quark model
estimates
in this paper as we believe that they are quite reliable for QCD-like
theories, see the discussion below.)

If we allow for a splitting in the technifermion masses the comparison
with experiment improves very
slightly. The values
of the effective lagrangian coefficients relevant for
the oblique corrections in the case of unequal masses are also given in
appendix E. Since $a_1$ is independent of the technifermion dynamically
generated masses anyway, the dependence is fully contained in $a_0$ (the
parameter $T$ of Peskin and Takeuchi\cite{PT}) and $a_8$ (the parameter
$U$). This is shown in figure~\ref{fig-2}.
We assume
that the splitting is the same for all doublets,
which is not necessarily true\footnote{In fact it can be argued that QCD
corrections may, in some cases\cite{holdom}, enhance techniquark masses.}.

If other representations of the $SU(2)_L\times SU(3)_c$ gauge group
are used, the oblique corrections have to be modified in the form
prescribed in section 8. Larger group theoretical factors lead
to larger oblique corrections and, from this point of view, the
restriction to weak doublets and colour singlets or triplets
is natural.

Let us close this section by justifying the use of chiral
quark model techniques, trying to assess the
errors involved, and at the same time emphasizing the importance
of having the scale dependence under control.
A parameter like $a_1$ (or $S$ in
the notation of Peskin and Takeuchi\cite{PT}) contains information about
the long-distance properties of a strongly coupled theory. In fact,
$a_1$ is nothing but the familiar $L_{10}$ parameter of the strong chiral
lagrangian of Gasser and Leutwyler\cite{GL} translated to the electroweak
sector. This strong interaction
parameter can be measured
and it is found to be $L_{10}= (-5.6\pm 0.3)\times 10^{-3}$ (at the
$\mu=M_\eta$ scale, which is just the conventional reference value
and plays no specific role in the Standard Model.)
This is almost twice the value predicted by the
chiral quark model\cite{AA,ERT} ($L_{10}= -1/32\pi^2$), which is
the estimate plotted in figure~\ref{fig-1}. Does this mean that the
chiral quark model grossly underestimates this observable?  Not at all.
Chiral perturbation theory predicts the running of $L_{10}$.
It is given by
\begin{equation}
L_{10}(\mu)=L_{10}(M_\eta)+\frac{1}{128\pi^2}\log\frac{\mu^2}{M_\eta^2}.
\end{equation}
According to our current understanding (see e.g. \cite{AET}),
the chiral quark model
gives the value of the chiral coefficients at the chiral symmetry
breaking scale ($4\pi f_\pi$ in QCD, $\Lambda_\chi$ in the electroweak
theory). Then the coefficient $ L_{10}$  (or $a_1$ for that matter)
predicted within the chiral quark model agrees with QCD at the 10\% level.

\fig{}{0cm}{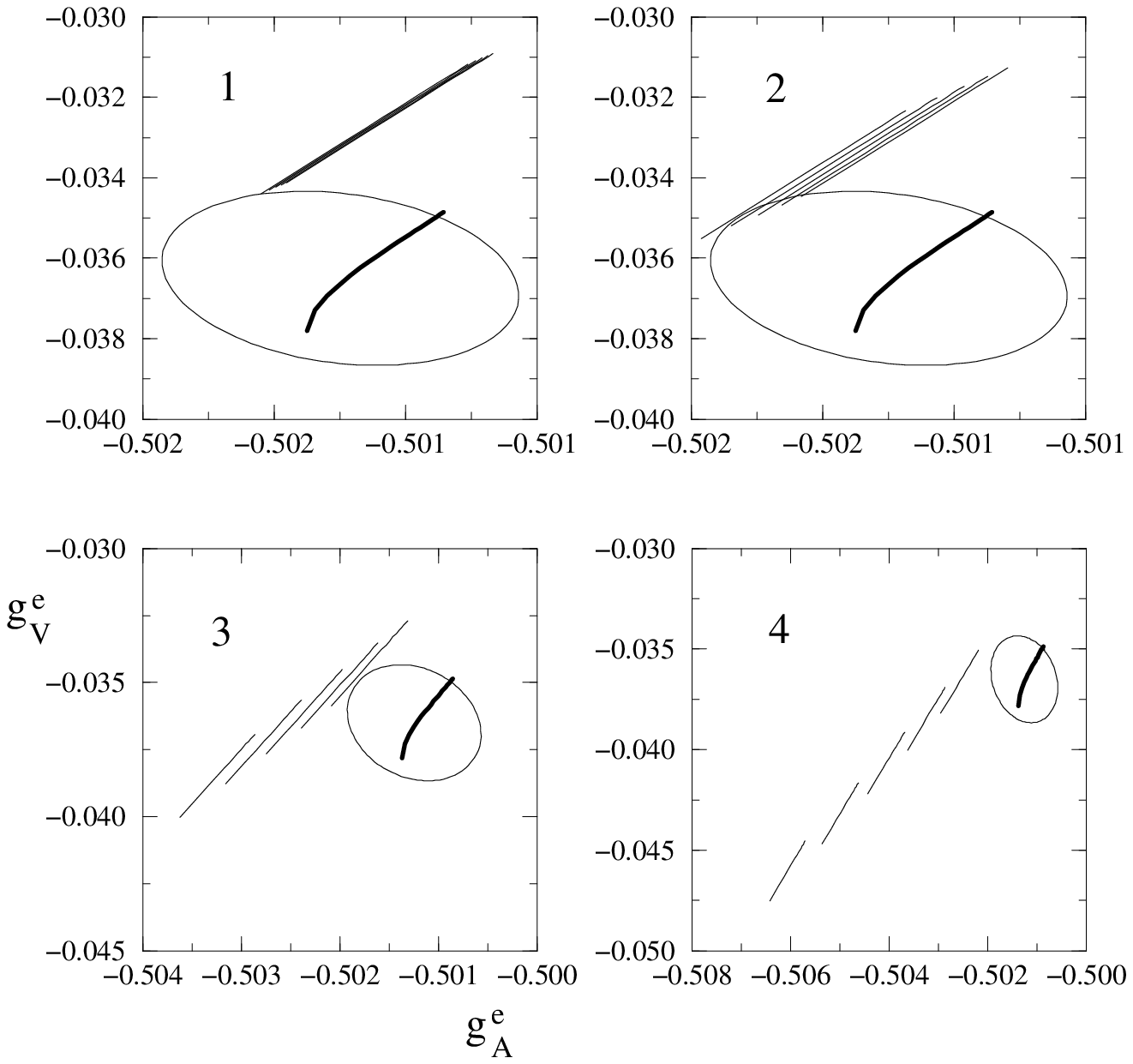}
{The effect of isospin breaking in the oblique corrections
in QCD-like technicolor theories.
The $1-\sigma$ region for the $g_A^e-g_V^e$
couplings and the SM prediction (for $m_t=175.6$ GeV, and
$70\le M_H\le 1000$ GeV) are shown. The different
straight lines correspond to setting the technifermion masses
in each doublet ($m_1$, $m_2$) to the value $m_2=$ 250,
300, 350, 400 and 450
GeV (larger masses are the ones deviating more from the
SM predictions), and $m_1=1.05 m_2$ (plot 1), $m_1=1.1 m_2$ (plot 2),
$m_1=1.2 m_2$
(plot 3), and $m_1=1.3 m_2$ (plot 4). The results are invariant under
the exchange of $m_1$ and $m_2$.
As in figure \ref{fig-1} the prediction of the effective
theory is the whole straight line and not any particular point on it, as we
move along the line by varying the unknown scale $\Lambda$. Clearly
isospin breakings larger than 20 \%
give very poor agreement with the data, even for low values of the
dynamically generated mass.}
{fig-2}
Let us now turn to the issue of vertex corrections in theories with
dynamical symmetry
breaking and the determination of the coefficients $\delta_i$ which
are, after all, the focal point of this work.

\section{New physics and four-fermion operators}
\label{new-phys}

In order to have a picture in our mind, let us
assume that at sufficiently high energies
the symmetry breaking sector can be described by
some renormalizable theory, perhaps a non-abelian gauge theory. By some
unspecified mechanism some
of the carriers of the new interaction acquire a mass. Let us generically
denote this mass by $M$.
One type of models that comes immediately to mind is the extended
technicolor scenario. $M$ would then be the mass of the ETC bosons. Let
us try, however,
not to adhere to any specific mechanism or model.

Below the scale $M$ we shall describe our underlying theory by four-fermion
operators. This is a convenient way of parametrizing the new physics below $%
M $ without needing to commit oneself to a particular model. Of course the
number of all possible four-fermion operators is enormous and one may think
that any predictive power is lost. This is not so because of two
reasons: a) The size of the
coefficients of the four fermion operators is not arbitrary. They are
constrained by the fact that at scale $M$ they are given by
\begin{equation}
-\xi_{\rm CG} \frac{G^2}{M^2}  \label{4qcoef}
\end{equation}
where $\xi_{\rm CG}$ is built out of Clebsch-Gordan factors and $G$ a gauge
coupling constant,
assumed perturbative of ${\cal O}(1)$ at the scale $M$.
The $\xi_{\rm CG}$ being
essentially group-theoretical factors are probably of similar
size for all three
generations, although not necessarily identical as this
would assume a particular style of embedding the different
generations into the large ETC (for instance) group. Notice that
for four-fermion operators of the form ${\bf J}\cdot
{\bf J}^\dagger$, where ${\bf J}$ is some fermion bilinear,
 $\xi_{\rm CG}$ has a well defined sign,
but this is not so for other operators.
b) It turns out that only a relatively small
number of combinations of these
coefficients do actually appear in physical observables at low energies.

Matching to the fundamental physical theory at $\mu=M$ fixes the
value of the coupling constants accompanying the four-fermion operators to
the value (\ref{4qcoef}). In addition contact terms, i.e.
non-zero values for the effective coupling constants $\delta_i$, are
generally speaking required
in order for the fundamental and four-fermion theories to match. These will
later evolve under the renormalization group due to the presence of the
four-fermion interactions. Because we expect that
$M\gg \Lambda_\chi$, the $%
\delta_i$ will be typically logarithmically enhanced. Notice that there
is no guarantee that this is the case for the
third generation, as we will later discuss. In this case the
TC and ETC dynamics would be tangled up (which for most models
is strongly disfavoured by the constraints on oblique corrections).
For the first and second generation, however, the logarithmic enhancement of
the $\delta_i$ is a potentially large
correction and it actually makes the treatment of a fundamental theory via
four-fermion operators largely independent of the particular details of
specific models, as we will see.

Let us now get back to four-fermion operators and proceed to a general
classification. A first observation is that, while in the bosonic sector
custodial symmetry is just broken by the small
$U(1)_Y$ gauge interactions, which
is relatively small, in the matter sector the breaking is not that small. We
thus have to assume that whatever underlying new physics is present at scale
$M$ it gives rise both to custodially preserving and custodially
non-preserving four-fermion operators with coefficients of similar strength.
Obvious requirements are hermiticity, Lorentz invariance and $SU(3)_c\times
SU(2)_L\times U(1)_Y$ symmetry. Neither $C$ nor $P$ invariance are imposed,
but invariance under $CP$ is assumed.

We are interested in $d=6$ four-fermion operators constructed with two
ordinary fermions (either leptons or quarks), denoted by $q_L$, $q_R$, and
two fermions $Q^{A}_L$, $Q^{A}_R$. Typically $A$ will be the technicolor
index and the $Q_L$, $Q_R$ will therefore be techniquarks and technileptons,
but we may be as well interested in the case where the $Q$ may be ordinary
fermions. In this case the index $A$ drops (in our subsequent formulae this
will correspond to taking $n_{TC}=1$). We shall not write the index $A$
hereafter for simplicity, but this degree of freedom is explicitly taken
into account in our results.

As we already mention we shall discuss in detail the case where
the additional fermions fall into ordinary representations of
$SU(2)_L\times SU(3)_c$ and will discuss other representations later.
The fields $Q_L$ will therefore transform as $SU(2)_L$ doublets and we
shall group the
right-handed fields $Q_R$ into doublets as well, but then include suitable
insertions of $\tau^3$ to consider custodially breaking operators. In order
to determine the low energy remnants of all these four-fermion operators
(i.e. the coefficients $\delta_i$) it is enough to know their couplings to $%
SU(2)_L$ and no further assumptions about their electric charges (or
hypercharges) are needed. Of course, since the $Q_L$, $Q_R$ couple to the
electroweak gauge bosons they must not lead to new anomalies. The simplest
possibility is to assume they reproduce the quantum numbers of one family of
quarks and leptons (that is, a total of four doublets $n_D=4$), but other
possibilities exist (for instance $n_D=1$ is also possible\cite{1doub},
although this model presents a global $SU(2)_L$ anomaly).

We shall first be concerned with the $Q_L$, $Q_R$ fields belonging to the
representation ${\bf 3}$ of $SU(3)_{c}$ and afterwards, focus in the simpler
case where the $Q_L$, $Q_R$ are colour singlet (technileptons). Coloured $%
Q_L$, $Q_R$ fermions can couple to ordinary quarks and leptons either
via the exchange of a colour singlet or of a colour octet. In addition the
exchanged particle can be either an $SU(2)_L$ triplet or a singlet, thus
leading to a large number of possible four-fermion operators. More important
for our purposes will be whether they flip or not the chirality. We use
Fierz rearrangements in order to write the four-fermion operators as product
of either two colour singlet or two colour octet currents. A complete list
is presented in table \ref{table-2} and table \ref{table-3}
for the chirality preserving and chirality flipping operators, respectively.

\begin{table}
\centering
\begin{tabular}{|c|c|}
\hline
$L^2=(\bar Q_{L} \ga_{\mu} Q_{L})(\bar q_{L}\ga^\mu q_{L})$ &  \\ \hline
$R^2=(\bar Q_{R}\ga_{\mu} Q_{R})(\bar q_{R}\ga^\mu q_{R})$ & $R_{3}R=(\bar
Q_{R}\ga_{\mu}\tau^3 Q_{R})(\bar q_{R}\ga^\mu q_{R})$ \\ \cline{2-2}
& $RR_{3}=(\bar Q_{R}\ga_{\mu} Q_{R})(\bar q_{R}\ga^\mu\tau^3 q_{R})$ \\
\cline{2-2}
& $R_{3}^2=(\bar Q_{R}\ga_{\mu}\tau^3 Q_{R})(\bar q_{R}\ga^\mu\tau^3 q_{R})$
\\ \hline\hline
$RL=(\bar Q_{R} \ga_{\mu} Q_{R})(\bar q_{L}\ga^\mu q_{L})$ & $R_{3}L=(\bar
Q_{R}\ga_{\mu} \tau^3 Q_{R})(\bar q_{L}\ga^\mu q_{L})$ \\ \hline
$LR=(\bar Q_{L}\ga_{\mu} Q_{L})(\bar q_{R}\ga^\mu q_{R})$ & $LR_{3}=(\bar
Q_{L}\ga_{\mu} Q_{L})(\bar q_{R}\ga^\mu \tau^3 q_{R})$ \\ \hline
$rl=(\bar Q_{R} \ga_{\mu}\vec\la Q_{R})\cdot (\bar q_{L}\ga^\mu\vec\la %
q_{L}) $ & $r_{3}l= (\bar Q_{R}\ga_{\mu}\vec\la \tau^3 Q_{R})\cdot (\bar
q_{L}\ga^\mu\vec\la q_{L})$ \\ \hline
$lr=(\bar Q_{L} \ga_{\mu}\vec\la Q_{L})\cdot (\bar q_{R}\ga^\mu\vec\la %
q_{R}) $ & $lr_{3}= (\bar Q_{L}\ga_{\mu}\vec\la Q_{L})\cdot (\bar q_{R}\ga%
^\mu\vec\la \tau^3 q_{R})$ \\ \hline\hline
$(\bar Q_{L}\ga_{\mu} q_{L})(\bar q_{L}\ga^\mu Q_{L})$ &  \\ \hline
$(\bar Q_{R}\ga_{\mu} q_{R})(\bar q_{R}\ga^\mu Q_{R}) $ & $(\bar Q_{R}\ga%
_{\mu}\tau^3 q_{R})(\bar q_{R}\ga^\mu Q_{R})+ (\bar Q_{R}\ga_{\mu}
q_{R})(\bar q_{R}\ga^\mu \tau^3 Q_{R})$ \\ \cline{2-2}
& $(\bar Q_{R}\ga_{\mu}\tau^3 q_{R})(\bar q_{R}\ga^\mu\tau^3 Q_{R})$ \\
\hline\hline
$(\bar Q_{L}^i\ga_{\mu} Q^j_{L} )(\bar q_{L}^j\ga^\mu q_{L}^i)$ &  \\ \hline
$(\bar Q_{R}^i\ga_{\mu} Q^j_{R} )(\bar q_{R}^j\ga^\mu q_{R}^i)$ &  \\ \hline
$(\bar Q_{L}^i\ga_{\mu} q_{L}^j)(\bar q_{L}^j\ga^\mu Q_{L}^i)$ &  \\ \hline
$(\bar Q_{R}^i\ga_{\mu}q_{R}^j) (\bar q_{R}^j\ga^\mu Q_{R}^i)$ & $(\bar
Q_{R}^i\ga_{\mu} q_{R}^j)(\bar q_{R}^j\ga^\mu [\tau^3 Q_{R}]^i)$ \\ \hline
\end{tabular}
\caption{Four-fermion operators which do not change the fermion chirality.
The first (second) column contains the custodially preserving (breaking)
operators.}
\label{table-2}
\end{table}
\begin{table}[ht]
\centering
\begin{tabular}{|c|c|}
\hline
\mbox{  $(\bar Q_{L}\ga^\mu q_{L})(\bar
q_{R}\ga_{\mu} Q_{R})$} &
\mbox{  $(\bar Q_{L}\ga^\mu q_{L})(\bar
q_{R}\ga_{\mu}\tau^3 Q_{R})$} \\ \hline
\mbox{ $(\bar
          q_{L}^i q_{R}^j)(\bar Q_{L}^k Q_{R}^l)\ep_{ik}\ep_{jl}$} &
\mbox{ $(\bar
          q_{L}^i  [\tau^3q_{R}]^j)(\bar Q_{L}^k
Q_{R}^l)\ep_{ik}\ep_{jl}$} \\ \hline
\mbox{ $(\bar
          q_{L}^i Q_{R}^j) (\bar Q_{L}^k q_{R}^l)\ep_{ik}\ep_{jl}$} &
\mbox{ $(\bar
          q_{L}^i Q_{R}^j) (\bar Q_{L}^k [\tau^3
q_{R}]^l)\ep_{ik}\ep_{jl}$} \\ \hline
\mbox{  $(\bar Q_{L}\ga^\mu\vec\la
q_{L})\cdot(\bar                  q_{R}\ga_{\mu}\vec\la Q_{R})$} &
\mbox{  $(\bar Q_{L}\ga^\mu\vec\la
q_{L})\cdot(\bar                         q_{R}\ga_{\mu}\vec\la\tau^3 Q_{R})$} \\ \hline
\mbox{ $(\bar
          q_{L}^i\vec\la q_{R}^j)\cdot( \bar Q_{L}^k\vec\la
Q_{R}^l)\ep_{ik}\ep_{jl}$} &
\mbox{ $(\bar
          q_{L}^i\vec\la [\tau^3q_{R}]^j) \cdot(\bar Q_{L}^k\vec\la
Q_{R}^l)\ep_{ik}\ep_{jl}$} \\ \hline
\mbox{ $(\bar
          q_{L}^i\vec\la Q_{R}^j)\cdot( \bar Q_{L}^k\vec\la
q_{R}^l)\ep_{ik}\ep_{jl}$} &
\mbox{ $(\bar
          q_{L}^i\vec\la Q_{R}^j)\cdot( \bar Q_{L}^k\vec\la
[\tau^3q_{R}]^l)\ep_{ik}\ep_{jl}$} \\ \hline
\end{tabular}
\caption{ Chirality-changing four-fermion operators. To each entry, the
corresponding hermitian conjugate operator should be added.
The left (right) column
contains custodially preserving (breaking) operators.}
\label{table-3}
\end{table}


\begin{table}
\centering
\begin{tabular}{|c|c|}
\hline
$l^2=(\bar Q_{L}\ga_{\mu}\vec\la Q_{L})\cdot(\bar q_{L}\ga^\mu\vec \la %
q_{L}) $ &  \\ \hline
$r^2=(\bar Q_{R}\ga_{\mu} \vec\la Q_{R})\cdot(\bar q_{R}\ga^\mu\vec\la %
q_{R}) $ & $r_{3}r=(\bar Q_{R}\ga_{\mu}\vec\la\tau^3 Q_{R})\cdot(\bar q_{R}%
\ga^\mu\vec\la q_{R})$ \\ \cline{2-2}
& $rr_{3}=(\bar Q_{R}\ga_{\mu}\vec\la Q_{R})\cdot(\bar q_{R}\ga^\mu\vec\la %
\tau^3 q_{R})$ \\ \cline{2-2}
& $r_{3}^2=(\bar Q_{R}\ga_{\mu}\vec\la\tau^3 Q_{R}) \cdot(\bar q_{R}\ga%
^\mu\vec\la\tau^3 q_{R})$ \\ \hline\hline
$\vec L^2=(\bar Q_{L}\ga_{\mu}\vec\tau Q_{L})\cdot (\bar q_{L}\ga%
^\mu\vec\tau q_{L})$ &  \\ \hline
$\vec R^2=(\bar Q_{R}\ga_{\mu}\vec\tau Q_{R})\cdot (\bar q_{R}\ga%
^\mu\vec\tau q_{R})$ &  \\ \hline
$\vec l^2=(\bar Q_{L}\ga_{\mu}\vec\la \vec\tau Q_{L})\cdot (\bar q_{L}\ga%
^\mu\vec\la \vec\tau q_{L})$ &  \\ \hline
$\vec r^2=(\bar Q_{R}\ga_{\mu}\vec\la \vec\tau Q_{R})\cdot (\bar q_{R}\ga%
^\mu\vec\la \vec\tau q_{R})$ &  \\ \hline
\end{tabular}
\caption{ New four-fermion operators of the form ${\bf J}\cdot{\bf j}
$ obtained after fierzing.
The left (right) column contains custodially preserving (breaking)
operators. In addition those written in the two upper blocks of table
\ref{table-2}
should also be considered. Together with the above they form a complete set
of chirality preserving operators.}
\label{table-4}
\end{table}

Note that the two upper blocks of table~\ref{table-2} contain operators of
the form ${\bf J}\cdot {\bf j}$, where (${\bf J}$) ${\bf j}$ stands for a
(heavy) fermion current with well defined colour and flavour numbers;
namely, belonging to an irreducible representation of $SU(3)_{c}$ and
$SU(2)_L$%
. In contrast, those in the two lower blocks are not of this form. In order
to make their physical content more transparent, we can perform a Fierz
transformation and replace the last nine operators (two lower blocks)
in table~\ref{table-2} by
those in table~\ref{table-4}.
These two basis are related by
\begin{eqnarray}
(\bar Q_L\ga_\mu q_L)(\bar q_L\ga^\mu Q_L)&=&\fth l^2+ \sixth L^2+\fth \vec
l~{}^2+\sixth \vec L^2 \\
(\bar Q_L^j\ga_\mu Q_L^i)(\bar q_L^i\ga^\mu q_L^j)&=& \half L^2+\half \LvLv
\\
(\bar Q_L^j\ga_\mu q_L^i)(\bar q_L^i\ga^\mu Q_L^j)&=& \half l^2 +\third L^2
\\
(\bar Q_R\ga_\mu q_R)(\bar q_R \ga^\mu Q_R)&=&\fth r^2 + \sixth R^2+\fth %
\rvrv+\sixth \RvRv \\
(\bar Q_R\ga_\mu q_R)(\bar q_R \ga^\mu \tau^3 Q_R)\kern1.5cm&&\nn \\
+ (\bar Q_R\ga_\mu\tau^3 q_R)(\bar q_R \ga^\mu Q_R)&=& \half rr_3+\third %
RR_3+\half r_3 r+ \third R_3R \\
(\bar Q_R\ga_\mu\tau^3 q_R)(\bar q_R \ga^\mu\tau^3 Q_R)&=& \fth r^2+\sixth %
R^2 -\fth \rvrv -\sixth \RvRv +\half r_3^2+\third R_3^2 \\
(\bar Q_R^j\ga_\mu Q_R^i)(\bar q_R^i\ga^\mu q_R^j)&=&\half R^2+ \half \RvRv
\\
(\bar Q_R^j\ga_\mu q_R^i)(\bar q_R^i\ga^\mu Q_R^j)&=&\half r^2+ \third R^2 \\
(\bar Q_R^j\ga_\mu q_R^i)(\bar q_R^i\ga^\mu [\tau^3 Q_R]^j)&=& \half r_3r+%
\third R_3R
\end{eqnarray}
for coloured techniquarks. Notice the appearance of
some minus signs due to the fierzing and that operators such as
$L^2$ (for instance) get contributions
from four fermions operators which do have a well defined sign as well
as from others which do not.

The use of this basis simplifies the calculations considerably as the Dirac
structure is simpler. Another
obvious advantage of this basis, which will become apparent only later, is
that it will make easier to consider the long distance contributions to the $%
\delta_i$, from the region of momenta $\mu < \Lambda_\chi$.

The  classification of the chirality preserving operator involving
technileptons is of course
simpler. Again we use Fierz rearrangements to write the operators
as ${\bf J}\cdot{\bf j}$. However, in this case only
a colour singlet ${\bf J}$ (and, thus, also a colour singlet ${\bf j}$) can
occur.
Hence, the  complete list can be obtained by crossing out from table
\ref{table-4} and from the first eight rows of table~\ref{table-2} the
operators involving
$\vec\la$. Namely, those designated by lower-case letters.
We are then left with the two operators $\vec L^2$, $\vec R^2$ from
table~\ref{table-4} and with the first six rows of table~\ref{table-2}:
$L^2$, $R^2$, $R_{3}R$, $RR_{3}$, $R_{3}^2$,
$RL$, $R_{3}L$, $LR$ and $LR_{3}$. If we choose to work instead with the
original basis of chirality preserving operators in
table~\ref{table-2}, we have to supplement these nine operators in the
first six
rows of the table with $(\bar Q_L\ga_\mu q_L)(\bar q_L\ga^\mu Q_L)$ and
$(\bar Q_R\ga_\mu q_R)(\bar q_R\ga^\mu Q_R)$, which are the only
independent ones  from the last seven rows. These two basis are related by
\begin{eqnarray}
(\bar Q_L\ga_\mu q_L)(\bar q_L\ga^\mu Q_L)&=&\half L^2+\half \LvLv \\
(\bar Q_R\ga_\mu q_R)(\bar q_R \ga^\mu Q_R)&=&\half R^2+\half \RvRv
\end{eqnarray}
for technileptons.

It should be borne in mind that Fierz transformations, as presented in the
above discussion, are strictly valid only in four dimensions. In $4
-2\epsilon$ dimensions for the identities to hold we need `evanescent'
operators\cite{evan}, which vanish in 4 dimensions. However the replacement
of some four-fermion operators in terms of others via the Fierz identities
is actually made inside a loop of technifermions and therefore a finite
contribution is generated. Thus the two basis will eventually be equivalent
up to terms of order
\begin{equation}
\frac{1}{16\pi^2}\frac{G^2}{M^2} m_Q^2  \label{fier}
\end{equation}
where $m_Q$ is the mass of the technifermion (this estimate will be obvious
only after the discussion in the next sections). In particular no logarithms
can appear in (\ref{fier}).

Let us now discuss how the appeareance of other representations
might enlarge the above classification.
We shall not be completely general here, but consider
only those operators that may actually contribute
to the observables we have been discussing (such as $g_V$ and $g_A$).
Furthermore, for reasons that shall be obvious in a moment,
we shall restrict ourselves to operators which are
$SU(2)_L\times SU(2)_R$ invariant.

The construction of the chirality conserving operators for fermions in
higher
dimensional representations of $SU(2)$ follows essentially the same pattern
presented in the appendix for doublet fields, except for the fact that
operators such as
\begin{equation}
(\bar Q_L \gamma_\mu q_L)(\bar q_L \gamma^\mu Q_L),\qquad
(\bar Q_L^i \gamma_\mu Q_L^j)(\bar q_L^j \gamma^\mu q_L^i),
\label{impossible}
\end{equation}
and their right-handed versions, which appear
on the right hand side of table 1,
are now obviously not acceptable since $Q_L$ and $q_L$ are in different
representations. Those operators,
restricting ourselves to color singlet bilinears (the only
ones giving a non-zero contribution to our observables) can be replaced
in the fundamental representation by
\begin{equation}
(\bar Q_L \gamma_\mu Q_L)(\bar q_L \gamma^\mu q_L),\qquad
(\bar Q_L \gamma_\mu\vec{\tau} Q_L)(\bar q_L \gamma^\mu\vec{\tau} q_L),
\label{possible}
\end{equation}
when we move to the ${\bf J}\cdot {\bf j}$ basis. Now it is
clear how to modify the above when using higher representations for
the $Q$ fields. The first one is already included in our set
of custodially preserving operators, while the second one has to be
modified to
\begin{equation}
{\vec L}^2\  \equiv \ (\bar Q_L \gamma_\mu \vec{T} Q_L)(\bar q_L
\gamma^\mu \vec{\tau}q_L), \label{higherrep}
\end{equation}
where $\vec{T}$ are the $SU(2)$ generators in the relevant
representation. In addition we have the right-handed
counterpart, of course.
We could in principle now proceed to construct custodially
violating operators by introducing suitable $T^3$ and $\tau^3$
matrices. Unfortunately, it is not possible to present a closed
set of operators of this type, as the number of independent operators
does obviously depend on the dimensionality of the representation.
For this reason we shall only consider custodially preserving
operators when moving to higher representations, namely
$L^2$, $R^2$, $RL$, $LR$, ${\vec L}^2$ and ${\vec R}^2$.

If we examine tables 1, 2 and 3 we will notice that both chirality
violating and chirality preserving operators appear.
It is clear that at the leading order in an expansion in external fermion
masses only the  chirality preserving operators (tables~\ref{table-2}
and~\ref{table-4}) are important, those operators
containing both a $q_L$ and a $q_R$  field will be further suppressed by
additional
powers of the masses of the fermions
and thus subleading. Furthermore, if we limit
our analysis to the study of the effective $W^\pm$ and $Z$ couplings, such
as $g_V$ and $g_A$,
as we do here,  chirality-flipping operators can contribute only through
a two-loop effect. Thus the contribution from the chirality
flipping operators contained in table 2 is suppressed both by an additional
$1/16\pi^2$ loop factor and by a $m_Q^2/M^2$ chirality factor. If
for the sake of the argument we take $m_Q$ to be 400 GeV, the correction
will be below or at the 10\% level for values of $M$ as low as 100 GeV.
This automatically eliminates from the game operators generated through the
exchange of a heavy scalar particle, but of course the presence
of light scalars, below the mentioned limit, renders their
neglection unjustified. It is not clear
where simple ETC models violate this limit (see e.g. \cite{mao}). We just
assume that all scalar particles can be made heavy enough.

Additional light scalars
may also appear as pseudo Goldstone bosons at the moment
the electroweak symmetry breaking occurs due to $\bar Q Q $
condensation. We had to assume somehow that their contribution
to the oblique correction was small (e.g. by avoiding their
proliferation and making them sufficiently heavy). They also
contribute to vertex corrections (and thus to the $\delta_i$), but
here their contribution is naturally suppresed. The coupling
of a pseudo Goldstone boson $\omega$ to ordinary fermions is of the form
\begin{equation}
\frac{1}{4\pi}\frac{m_Q^2}{M^2} \omega \bar{q}_L q_R, \label{psse}
\end{equation}
thus their contribution to the $\delta_i$ will be or order
\begin{equation}
g\frac{G^4}{(16\pi^2)^2}(\frac{m_Q^2}{M^2})^2\log\frac{\Lambda_\chi^2}{m_\omega^2}.
\label{contrismall}
\end{equation}
Using the same reference values as above
a pseudo Goldstone boson of 100 GeV can be neglected.

If the operators contained in table 2 are not relevant for
the $W^\pm$ and $Z$ couplings, what are they important for?
After electroweak breaking (due to the strong technicolor
forces or any other mechanism) a condensate $\langle \bar
Q Q\rangle $ emerges. The chirality flipping operators
are then responsible for generating a mass term for ordinary
quarks and leptons. Their low energy effects are contained
in the only $d=3$ operator appearing in the matter
sector, discussed in section 2. We thus see that
the four fermion approach allows for a nice separation
between the operators responsible for mass generation
and those that may eventually lead to observable consequences
in the $W^\pm$ and $Z$ couplings. One may even entertain
the possibility that the relevant scale is, for some
reason, different for both sets of operators (or, at least,
for some of them). It could, at least in principle,  be the case that
scalar exchange enhances the effect of chirality flipping
operators, allowing for large masses for the third
generation, without giving unacceptably large contributions
to the $Z$ effective coupling. Whether one is able
to find a satisfactory fundamental theory where
this is the case is another matter, but the four-fermion approach
allows, at least, to pose the problem.

We shall now proceed to determine the constants $\delta_i$
appearing in the effective lagrangian after
integration of the heavy degrees of freedom.
For the sake of the discussion we shall assume
hereafter that technifermions are
degenerate in mass and set their masses equal to $m_Q$. The general
case is discussed in appendix E.

\section{Matching to a fundamental theory}

At the scale $\mu=M$ we integrate out the heavier degrees of freedom by
matching the renormalised Green functions computed in the underlying
fundamental theory to a four-fermion interaction. This matching leads to the
values (\ref{4qcoef}) for the coefficients of the four-fermion operators as
well as to a purely short distance contribution for the $\de_i$, which shall
be denoted by $\tilde\delta_i$. The matching procedure is indicated in
figure~\ref{fig-5}.
\fig{}{0cm}{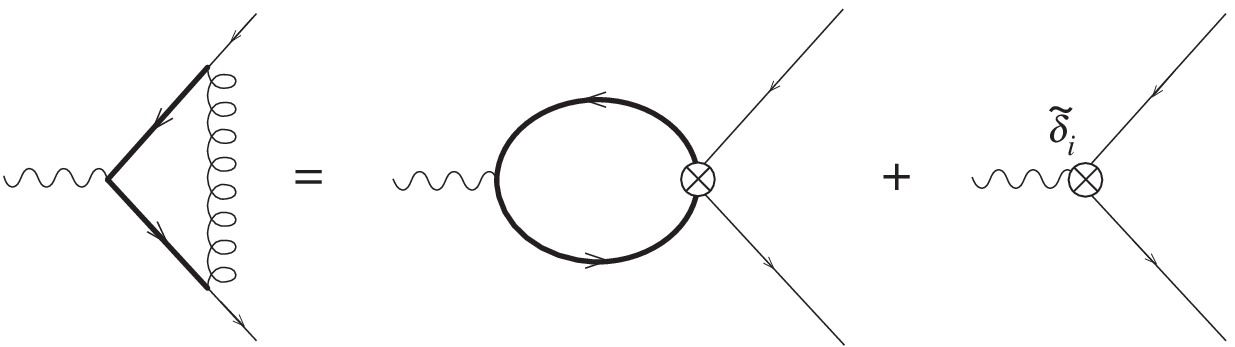}{The matching at the scale $\mu=M$.}{fig-5}
It is perhaps useful to think of the $\tilde\delta_i$ as the value
that the coefficients of the effective lagrangian take at the matching scale,
as they contain the information on modes of frequencies $\mu>M$. The $%
\tilde\delta_i$ will be, in general, divergent, i.e. they will have a pole
in $1/\epsilon$. Let us see how to obtain these coefficients $\tilde\delta_i$
in a particular case.

As discussed in the previous section we understand that at very high
energies our theory is described by a gauge theory. Therefore we have to
add to the
Standard Model lagrangian (already extended with technifermions)
the following pieces
\begin{equation}
-\frac{1}{4}E_{\mu\nu} E^{\mu\nu}
-\frac{1}{2} M^2 E_\mu E^\mu + G \bar Q \gamma^\mu E_\mu q + {\rm h. c.}.
\label{Eboson}
\end{equation}
The $E_\mu$ vector boson (of mass $M$) acts in a large flavour group space
which mixes ordinary fermions with heavy ones.
(The notation in (\ref{Eboson}) is somewhat symbolic as we are not
implying that the theory is vector-like, in fact we do not assume
anything at all about it.)

At energies $\mu < M$ we can describe the contribution from this sector to
the effective lagrangian coefficients either using the degrees of freedom
present in (\ref{Eboson}) or via the corresponding four quark operator and a
non-zero value for the $\tilde\delta_i$ coefficients. Demanding that both
descriptions reproduce the same renormalised $ffW$ vertex fixes the
value of the $\tilde\delta_i$.

Let us see this explicitly in the case where the intermediate vector boson
$E_\mu$ is a $SU(3)_c\times SU(2)_L$ singlet. For the sake of
simplicity, we take the third term in~(\ref{Eboson}) to be
\begin{equation}
     G \bar Q_{L} \gamma^\mu E_\mu q_{L} .
     \label{Eboson'}
\end{equation}
At energies below $M$, the relevant four quark operator is then
\begin{equation}
     -{G^2\over M^2} (\bar Q_{L}\ga^\mu q_{L})(\bar q_{L}\ga_\mu Q_{L}) .
     \label{4q-1}
\end{equation}
In the limit of degenerate techniquark masses, it is quite clear that
only $\tilde\de_{1}$ can be different from zero. Thus, one does not
need to worry about matching quark self-energies. Concerning the
vertex (figure~\ref{fig-5}), we have to impose eq.~(\ref{match-2}),
where now
\begin{equation}
     \De\Gamma\equiv \Gamma_{E}-\Gamma_{4Q} .
     \label{match-1}
\end{equation}
Namely, $\De\Gamma$ is the difference between the vertex computed
using~(\ref{Eboson}) and
the same quantity computed using the four quark operators as well as
non zero $\tilde\de_{i}$ coefficients (recall that the hat
in~(\ref{match-2}) denotes
renormalised quantities). A calculation analogous to
that of section~\ref{31.7-1} (now the leading terms in $1/M^2$ are
retained) leads to  \begin{equation}
     \tilde\de_{1}=-{G^2\over8\pi^2}{m_{Q}^2\over
     M^2}{1\over\hat\ep}.
     \label{de-1}
 \end{equation}

\section{Integrating out heavy fermions}

As we move down in energies we can integrate lower and lower frequencies
with the help of the four-fermion operators (which do accurately describe
physics below $M$). This modifies the value of the $\delta_i$
\begin{equation}
\delta_i(\mu)= \tilde\delta_i+ \Delta \delta_i(\mu/M),\qquad \mu<M.
\label{short}
\end{equation}
The quantity $\Delta \delta_i(\mu/M)$ can be computed in perturbation theory
down to the scale $\Lambda_\chi$ where the residual interactions labelled by
the index $A$ becomes strong and confine the technifermions. The leading
contribution is given by a loop of technifermions.

To determine such contribution it is necessary to demand that the
renormalised Green functions match when computed using explicitly the degrees
of freedom $Q_L$, $Q_R$ and when their effect is described via the effective
lagrangian coefficients $\delta_i$. The matching procedure is illustrated in
figure~\ref{fig-6}.
\fig{}{0cm}{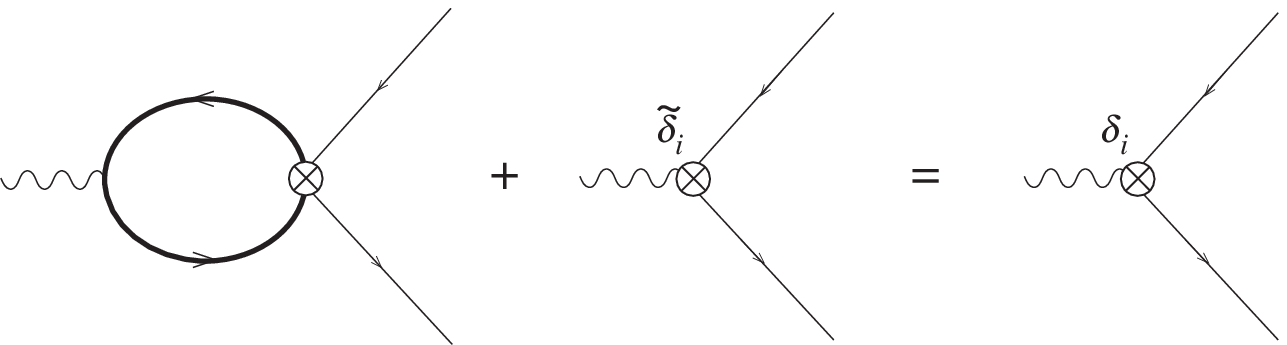}{Matching at the scale $\mu=\Lambda_\chi$.}{fig-6}
The scale $\mu$ of the matching must be such that $\mu<M$, but
such that $\mu>\Lambda_\chi$, where perturbation theory in the technicolour
coupling constant starts being questionable.

The result of the calculation in the case of degenerate masses is
\begin{equation}
\Delta\delta_i(\mu/M) =-\bar\delta_i\left(1 - \hat\ep
\log{\mu^2\over M^2}\right),  \label{logg}
\end{equation}
where we have kept the logarithmically enhanced contribution only and have
neglected any other possible constant pieces. $\bar\delta_i$ is the singular
part of $\tilde\delta_i$. The finite parts of $\tilde\delta_i$ are clearly
very model dependent (cfr. for instance the previous discussion on
evanescent operators) and we cannot possibly take them into account in a
general analysis. Accordingly, we ignore all other terms in (\ref{logg}) as well
as those finite pieces generated through the fierzing procedure (see
discussion in previous section). Keeping
the logarithmically enhanced terms
therefore sets the level of accuracy of our calculation. We will call (\ref
{short}) the short-distance contribution to the coefficient $\delta_i$.
General formulae for the case where the two technifermions are not
degenerate in masses can be found in appendix E.

Notice that the final short distance contribution to the $\delta_i$ is
ultraviolet finite, as it should. The divergences in $\tilde\delta_i$ are
exactly matched by those in $\Delta\delta_i$. The pole in $\tilde\delta_i$
combined with singularity in $\Delta\delta_i$ provides a finite contribution.

There is another potential source of corrections to the $\delta_i$ stemming
from the renormalization of the four fermion coupling constant $G^2/M^2$
(similar to the renormalization of the Fermi constant in the electroweak
theory due to gluon exchange). This effect is however subleading here. The
reason is that we are considering technigluon exchange only for four-fermion
operators of the form ${\bf J}\cdot {\bf j}$, where, again, ${\bf j}$
(${\bf J}$)
stands for a (heavy) fermion current (which give the leading contribution,
as discussed). The fields carrying technicolour have the same handedness and
thus there is no multiplicative renormalization and the effect is absent.

Of course in addition to the short distance contribution there is a
long-distance contribution from the region of integration of momenta $%
\mu<\Lambda_\chi$. Perturbation theory in the technicolour coupling constant
is questionable and we have to resort to other methods to determine the
value of the $\delta_i$ at the $Z$ mass.

There are two possible ways of doing so. One is simply to mimic the
constituent chiral quark model of QCD. There one loop of chiral quarks with
momentum running between the scale of chiral symmetry breaking and the scale
of the constituent mass of the quark, which acts as infrared cut-off,
provide the bulk of the contribution\cite{ERT,AET} to $f_\pi $, which is the
equivalent of $v$. Making the necessary translations we can write
for QCD-like theories
\begin{equation}
v^2\simeq n_{TC} n_D\frac{m_Q^2}{4\pi^2}\log\frac{\Lambda^2_\chi}{m_Q^2}.
\label{v2}
\end{equation}

Alternatively, we can use chiral lagrangian techniques\cite{HG} to
write a low-energy bosonized version of the technifermion bilinears $\bar
Q_L \Gamma Q_L$ and $\bar Q_R \Gamma Q_R$
using the chiral currents ${\bf J}%
_L$ and ${\bf J}_R$. The translation is
\begin{equation}
\bar Q_L \gamma^\mu Q_L \to \frac{v^2}{2} {\rm tr} U^\dagger\ri D_\mu U
\end{equation}
\begin{equation}
\bar Q_L \gamma^\mu \tau^i Q_L \to \frac{v^2}{2} {\rm tr} U^\dagger
\tau^i\ri D_\mu U
\end{equation}
\begin{equation}
\bar Q_R \gamma^\mu Q_R \to\frac{v^2}{2} {\rm tr} U\ri D_\mu U^\dagger
\end{equation}
\begin{equation}
\bar Q_R \gamma^\mu \tau^i Q_R \to\frac{v^2}{2} {\rm tr} U \tau^i\ri D_\mu
U^\dagger
\end{equation}
Other currents do not contribute to the effective coefficients. Both methods
agree.

Finally, we collect all
contributions to the coefficients $\delta_i$ of the
effective lagrangian. For fields in the usual representations
of the gauge group
\begin{eqnarray}
\delta_1&=& a_{\vec L^2}\frac{G^2}{M^2} (v^2 + n_{TC} n_D
\frac{m_Q^2}{4\pi^2} \log\frac{M^2}{\Lambda^2_\chi})
-\frac{1}{16\pi^2}\frac{y_u^2+y_d^2}{4}(\frac{1}{\hat\epsilon}
-\log\frac{\Lambda^2}{\mu^2})\\
\delta_2&=& (a_{\vec R^2} +\frac{1}{2} a_{R_3^2}) \frac{G^2}{M^2} (v^2 +
n_{TC} n_D \frac{m_Q^2}{4\pi^2} \log\frac{M^2}{\Lambda^2_\chi})
-\frac{1}{16\pi^2}\frac{(y_u+y_d)^2}{8}(\frac{1}{\hat\epsilon}
-\log\frac{\Lambda^2}{\mu^2})\\
\delta_3&=& \frac{1}{2}a_{R_3 L} \frac{G^2}{M^2} (v^2 + n_{TC} n_D \frac{%
m_Q^2}{4\pi^2} \log\frac{M^2}{\Lambda^2_\chi})
+\frac{1}{16\pi^2}\frac{y_u^2-y_d^2}{4}(\frac{1}{\hat\epsilon}
-\log\frac{\Lambda^2}{\mu^2})\\
\delta_4&=& 0\\
\delta_5&=& \frac{1}{2}a_{R_3 R} \frac{G^2}{M^2} (v^2 + n_{TC} n_D \frac{%
m_Q^2}{4\pi^2} \log\frac{M^2}{\Lambda^2_\chi})
-\frac{1}{16\pi^2}\frac{y_u^2-y_d^2}{4}(\frac{1}{\hat\epsilon}
-\log\frac{\Lambda^2}{\mu^2})\\
\delta_6&=& \frac{1}{2} a_{R_3^2} \frac{G^2}{M^2} (v^2 + n_{TC} n_D \frac{%
m_Q^2}{4\pi^2} \log\frac{M^2}{\Lambda^2_\chi})
-\frac{1}{16\pi^2}\frac{(y_u-y_d)^2}{4}(\frac{1}{\hat\epsilon}
-\log\frac{\Lambda^2}{\mu^2}),
\end{eqnarray}
while in the case of higher representations, where only
custodially preserving operators have been considered, only
$\delta_1$ and $\delta_2$ get non-zero values (through
$a_{{\vec L}^2}$ and $a_{{\vec R}^2}$). The long distance
contribution is, obviously, universal (see section 2),
while we have to modify the short distance contribution
by replacing the Casimir of the fundamental representation
of $SU(2)$ for the appropriate one ($1/2 \to c(R)$),
the number of doublets by the multiplicity of the given
representation, and $n_c$ by the appropriate dimensionality
of the $SU(3)_c$ representation to which the $Q$ fields
belong.

These expressions require several comments. First of all, they
contain the same (universal) divergences as their
counterparts in the minimal Standard Model. The scale $\Lambda$
should, in principle, correspond to the matching scale $\Lambda_\chi$,
where the low-energy non-linear effective theory takes over.
However, we write an arbitrary scale just to remind us
that the finite part accompanying
the log is regulator dependent and cannot
be
determined within the effective theory. Recall that the
leading ${\cal O}(n_{TC}n_D)$
term is finite and unambiguous, and that the ambiguity lies
in the formally subleading term (which, however, due to the log is
numerically quite important).
Furthermore only logarithmically enhanced terms
are included in the above expressions. Finally one should bear in mind
that the chiral quark model techniques that we have used are
accurate only in the large $n_{TC}$ expansion (actually $n_{TC}n_D$ here).
The same comments apply of course to the oblique coefficients $a_i$
presented in the appendix.

The quantities $a_{\vec L^2}$, $a_{\vec R^2}$, $a_{R_3^2}$,
$a_{R_3 L}$ and $a_{R_3 R}$ are the coefficients of the
four-fermion operators indicated by the sub-index
(a combination of Clebsch-Gordan
and fierzing factors). They depend on the specific model.
As discussed in previous sections these coefficients can be of either
sign. This
observation is important because it shows that the contribution to the
effective coefficients has no definite sign\cite{noncomm} indeed.
It is nice that there is almost a one-to-one correspondence between
the effective lagrangian coefficients (all of them measurable, at least
in principle) and four-fermion coefficients.

Apart from these four-fermion coefficients, the $\delta_i$ depend
on a number of quantities ($v$, $m_Q$, $\Lambda_\chi$, $G$ and $M$).
Let us first discuss those related to the electroweak symmetry breaking,
($m_Q$ and $\Lambda_\chi$)
and postpone the considerations on $M$ to the next section ($G$ will be
assumed to be of ${\cal O}(1)$). $v$ is of course the Fermi scale and hence
not an unknown at all ($v\simeq 250$ GeV).
The value of $m_Q$ can be estimated
from (\ref{v2}) since $v^2$ is known and $\Lambda_\chi$, for
QCD-like technicolor theories is $\sim 4\pi v$. Solving for $m_Q$
one finds that if $n_D=4$, $m_Q\simeq v$, while if $n_D=1$,
$m_Q\simeq 2.5 v$. Notice that $m_Q$ and $v$ depend differently
on $n_{TC}$ so it is not correct to simply assume $m_Q\simeq v$.
In theories where the technicolor $\beta$ function is small
(and it is pretty small if  $n_D=4$ and $n_{TC}=2$) the characteristic scale
of the breaking is pushed upwards, so we expect $\Lambda_\chi \gg
 4\pi v$. This brings $m_Q$ somewhat downwards, but the
decrease is only logarithmic. We shall therefore take $m_Q$
to be in the range 250 to 450 GeV. We shall allow for a mass splitting
within the doublets too. The splitting within each doublet cannot be too
large, as figure \ref{fig-2} shows. For simplicity we
shall assume an equal splitting of
masses for all doublets.

\section{Results and discussion}

Let us first summarize our results so far.
The values of the effective lagrangian coefficients encode
the information about the symmetry breaking sector
that is (and will be in the near future) experimentally
accessible. The $\delta_i$ are therefore the
counterpart of the oblique corrections coefficients
$a_i$ and they have to be taken together in
precision analysis of the Standard Model, even if
they are numerically less significant.

These effective coefficients
apply to $Z$-physics at LEP, top production at the Next
Linear Collider, measurements of the top decay at CDF,
or indeed any other process involving the third generation
(where their effect is largest), provided the energy involved
is below  $4\pi v$,  the limit of applicability of chiral techniques.
(Of course  chiral effective lagrangian techniques
fails well below $4\pi v$  if a resonance
is present in a given channel, see also \cite{gd}.)

In the Standard model the $\delta_i$ are useful to keep track of the
$\log M_H$ dependence in all processes involving either
neutral or charged currents.  They also provide an economical description
of the symmetry breaking sector, in the sense that they contain the relevant
information in the low-energy regime, the only one testable at present.
Beyond the Standard model the new physics contributions
is parametrized by four-fermion operators. By choosing
the number of doublets, $m_Q$, $M$, and $\Lambda_\chi$ suitably, we
are in fact describing in a single shot a variety of theories: extended
technicolor (commuting and non-commuting),
walking technicolor\cite{walking} or top-assisted technicolor, provided
that all remaining scalars and pseudo-Goldstone bosons are sufficiently
heavy.

The accuracy of the calculation
is limited by a number of approximations we have been
forced to make and which have been discussed at length
in previous sections. In practice we retain only terms which are
logarithmically enhanced when running from $M$ to $m_Q$, including
the long distance part, below $\Lambda_\chi$.
The effective lagrangian coefficients $\delta_i$ are all finite at the scale
$\Lambda_\chi$, the lower limit of applicability of
perturbation theory. Below that scale they run following the
renormalization group equations of the non-linear theory and
new divergences have to be subtracted\footnote{The divergent contribution
coming
from the Standard Model $\delta_i$'s has to be removed, though, as discussed
in section 5, so the difference is finite and would be fully predictable,
had we good theoretical control on the subleading corrections. At present
only the ${\cal O}(n_{TC}n_D)$ contribution is under reasonable control.}.
These coefficients contain finally
the contribution from scales $M> \mu >m_Q$, the dynamically generated
mass of the technifermion (expected to
be
of ${\cal O}(\Lambda_{TC})$.
In view of the theoretical uncertainties, to restrict
oneself to logarithmically enhanced terms is a very reasonable approximation
which should capture the bulk of the contribution.

Let us now proceed to a more detailed discussion of
the implications of our analysis. Let us begin by discussing
the value that we should take for $M$, the mass scale normalizing
four-fermion operators.
Fermion condensation gives a mass
to ordinary fermions via chirality-flipping operators of order
\begin{equation}
m_f\simeq\frac{G^2}{M^2}\langle \bar Q Q\rangle , \label{mfer}
\end{equation}
through the operators listed in table \ref{table-3}.
A chiral quark model calculation shows that
\begin{equation}
 \langle \bar Q Q\rangle \simeq v^2 m_Q.
\end{equation}
Thus, while $\langle \bar Q Q\rangle $ is
universal, there is an inverse relation between $M^2$ and $m_f$.
In QCD-like theories this leads to the following rough estimates
for the mass $M$ (the subindex refers to the fermion
which has been used in the l.h.s. of (\ref{mfer}))
\begin{equation}
M_e\sim 150 {\rm TeV},\qquad M_\mu\sim  10 {\rm TeV},\qquad
M_b\sim 3 {\rm TeV}.
\end{equation}
If taken at face value, the scale for $M_b$ is too low,
even the one for $M_\mu$ may already conflict with current
bounds on FCNC, unless they are suppressed by some other mechanism
in a natural way. Worse, the top mass cannot be reasonably
reproduced by this mechanism. This well-known problem
can be partly alleviated
in theories where technicolor walks or invoking
top-colour or a similar
mechanism \cite{topcolor}). Then $M$ can be made larger and $m_Q$, as
discussed, somewhat smaller.
For theories which are
not vector-like the
above estimates become
a lot less reliable.

However one should not forget that
none of the four-fermion operators
playing a role in the vertex effective couplings participates at all in the
fermion mass determination.
In principle we can then entertain the
possibility that the relevant mass scale for the latter should be
lower (perhaps because they get a contribution through scalar exchange,
as some of them can be generated this way).
Even in this case it seems just natural that $M_b$ (the scale
normalizing chirality preserving operators for the third generation, that
is) is low and not too different from $\Lambda_\chi$.
Thus the logarithmic
enhancement is pretty much absent in this case and some of the
approximations made become quite questionable in this case. (Although even
for the $b$ couplings there is still a relatively large contribution to the
$\delta_i$'s coming from long distance contributions.)
Put in another words, unless an additional mechanism is invoked,
it is not really possible to make definite
estimates for the $b$-effective couplings without getting into
the details of the underlying theory. The flavour
dynamics and electroweak breaking are completely
entangled in this case. If one only retains the long distance
part (which is what we have done in practice) we can, at
best, make order-of-magnitude estimates. However, what is remarkable
in a way is that this does not happen for the first and second
generation vertex corrections. The effect of flavour dynamics
can then be encoded in a small number of coefficients.

We shall now discuss in some detail the numerical consequences of our
assumptions. We shall assume the above values for the mass scale $M$;
in other words, we shall place ourselves in the most disfavourable
situation.
We shall only present results for QCD-like
theories and $n_D=4$ exclusively.
For other theories the appropriate results can be very
easily obtained from our formulae.
For the coefficients $a_{\vec L^2}$,  $a_{R_3R}$, $a_{R_3 L}$, etc.
we shall use the range of variation [-2, 2] (since
they are expected to be of ${\cal O}(1)$). Of course larger values
of the scale, $M$, would simply translate into smaller
values for those coefficients, so the results can be easily scaled down.

Figure~\ref{fig-7} shows the $g_A^e, g_V^e$ electron
effective couplings when vertex corrections are included and allowed
to vary within the stated limits. To avoid clutter, the top mass
is taken to the central value 175.6 GeV. The Standard Model
prediction is shown as a function of the Higgs mass. The dotted lines
in figure \ref{fig-7} correspond to considering the oblique corrections
only. Vertex corrections change these results and, depending on the
values of the four-fermion operator coefficients, the prediction can
take any value in the strip limited by the two solid lines (as usual
we have no specific prediction in the direction along the strip due
to the dependence on $\Lambda$, inherited from the non-renormalizable
character of the effective theory). A generic modification of
the electron couplings is of ${\cal O}(10^{-5})$, small but much larger
than in the Standard Model and, depending on its sign, may help
to bring a better agreement with the central value.

The modifications are more dramatic in the case of the second generation,
for the muon, for instance. Now, we expect changes in the $\delta_i$'s
and, eventually, in the effective couplings of ${\cal O}(10^{-3})$
These modifications are just at the limit of being observable. They
could even modify the relation between $M_W$ and $G_\mu$ (i.e.
$\Delta r$).

Figure~\ref{fig-8} shows a similar plot for the bottom effective couplings
$g_A^b, g_V^b$.
It is obvious that taking generic values for the four-fermion
operators (of ${\cal O}(1)$) leads to enormous modifications
in the effective couplings, unacceptably large in fact. The corrections
become more manageable if we allow for a smaller variation of the
four-fermion operator coefficients (in the range [-0.1,0.1]).
This suggests that the natural order of magnitude for the
mass $M_b$ is $\sim 10$ TeV, at least for chirality preserving
operators.
As we have discussed the corrections can be of
either sign.

\fig{}{0cm}{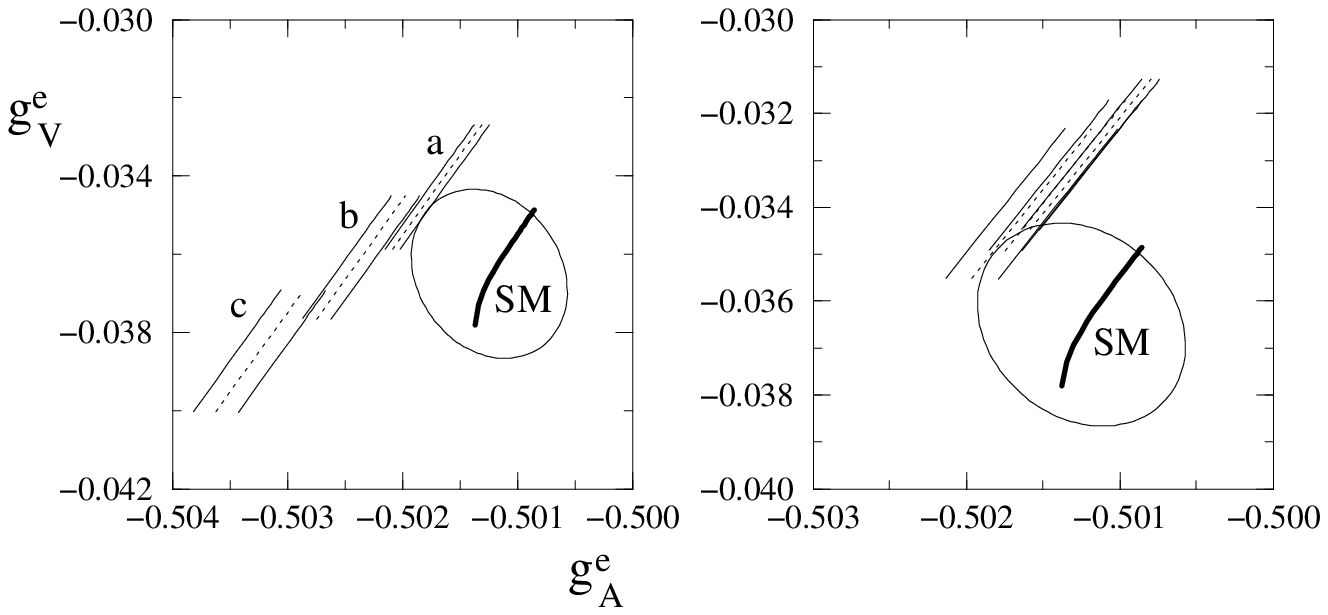}{Oblique and vertex corrections for the electron
effective couplings. The elipse indicate the 1-$\sigma$ experimental region. 
Three values of the effective mass $m_2$ are
considered: 250 (a), 350 (b) and 450 GeV (c), and two splittings: 
10\% (right) and 20\% (left). The dotted lines correspond to including 
the oblique corrections only. The coefficients of the four-fermion 
operators vary in the range [-2,2] and this spans the region between 
the two solid lines. The Standard Model prediction (thick solid line) 
is shown for $m_t=175.6$ GeV and $70\le M_H\le 1500$ GeV.}{fig-7}

\bigskip

\fig{}{0cm}{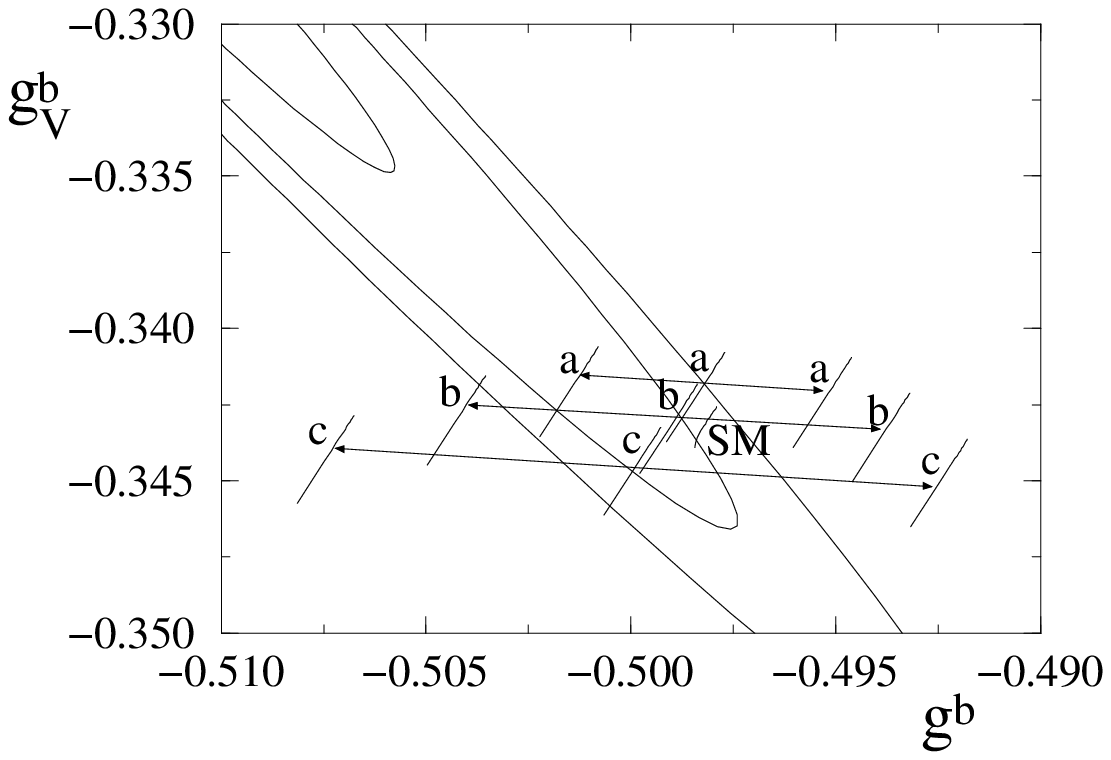}
{Bottom effective couplings compared to the SM prediction for $m_t=175.6$ 
as a function of the Higgs mass (in the range [70,1500] GeV). The 
elipses indicate 1, 2, and 3-$\sigma$ experimental regions.
The dynamically generated masses are 250 (a), 350 (b) and 400 GeV (c) and 
we show a 20\% splitting between the masses in the heavy doublet. 
The degenerate case does not present quantitative differencies if 
we consider the experimental errors. The central lines correspond to 
including only the oblique corrections. When we include 
the vertex corrections (depending on the size of the four-fermion 
coefficients) we predict the regions between lines indicated by the arrows. 
The four-fermion coefficients in this case take values in the range 
[-0.1,0.1].}{fig-8}

One could, at least in the case of degenerate masses, translate
the experimental constraints on the $\delta_i$ (recall that their
experimental determination requires a combination of charged and neutral
processes, since there are six of them) to the coefficients of
the four-fermion operators. Doing so would provide us with a four-fermion
effective
theory that would exactly reproduce all the available data. It is obvious
however that the result would not be very satisfactory. While the outcome
would, most likely, be coefficients of ${\cal O}(1)$ for the electron
couplings, they would have to be of ${\cal O}(10^{-1})$, perhaps
smaller for the bottom. Worse, the same masses we have used
lead to unacceptably low values for the top mass (\ref{mfer}).
Allowing for a different scale in the chirality flipping operators would
permit
a large top mass without affecting the effective couplings. Taking this as a
tentative possibility we can
pose the following problem: measure the effective couplings $\delta_i$
for all three generations and determine the values of the four-fermion
operator coefficients and the characteristic mass scale that
fits the data best. In the degenerate mass limit we have a total
of 8 unknowns (5 of them coefficients, expected to be of ${\cal O}(1)$)
and 18 experimental values (three sets of the $\delta_i$).  A similar
exercise could be attempted in the chirality flipping sector. If
the solution to this exercise turned out to be mathematically consistent
(within the experimental errors) it would be extremely interesting.
A negative result would certainly rule out this approach.
Notice that dynamical symmetry breaking predicts the
pattern $\delta_i\sim m_f$, while in the Standard Model
$\delta_i\sim m_f^2$.

We should end with some words of self-criticism. It may seem
that the previous discussion is not too conclusive and that
we have managed only to rephrase some of the
long-standing problems in the symmetry breaking sector.
However, the {\it raison d'\^etre}
of the present paper is not really to propose a solution
to these problems, but rather to establish a theoretical
framework to treat them systematically. Experience from the past shows
that often the effects of new physics are magnified and thus models are
ruled out on this basis, only to find out that a careful and rigorous
analysis leaves some room for them. We believe that this may be the case
in dynamical symmetry breaking models and we believe too that only through a
detailed and careful
comparison with the experimental data will progress take place.

The effective lagrangian
provides the tools to look for an `existence proof' (or otherwise) of
a phenomenologically viable, mathematically consistent
dynamical symmetry breaking model. We hope that there is any time soon
sufficient experimental data to attempt to determine the four-fermion
coefficients, at lest approximately.

\bigskip

{\bf Acknowledgements}

We would like to thank M.J. Herrero, M. Martinez, J. Matias, S. Peris, J.
Taron and F. Teubert for discussions. D.E. wishes to thank the hospitality
of the SLAC Theory Group where this work was finished. J.M. acknowledges a
fellowship from Generalitat de Catalunya, grant 1998FI-00614. This work has
been partially supported by CICYT grant AEN950590-0695 and CIRIT contract
GRQ93-1047.

\bigskip

\newpage

\part*{Appendices}

\appendix

\section{{\ $d=4$} operators}

\label{app-3.6a}

The procedure we have followed to obtain~(\ref{2.6b}--\ref{2.6d}) is very
simple. We have to look for operators of the form $\bar\psi\Gamma\psi$,
where $\psi=q_{L},\;q_{R}$ and $\Gamma$ contains a covariant derivative, $%
D_{\mu}$, and an arbitrary number of $U$ matrices. These operators must be
gauge invariant so not any form of $\Gamma$ is possible.
Moreover, we can drop total derivatives and, since $U$ is unitary, we have
the following relation
\begin{equation}
D_{\mu}U=-U(D_{\mu} U)^\dagger U.
\end{equation}
Apart from the obvious structure $D_{\mu} U$ which transform as $U$ does,
we immediately realise that the particular form of $G_{R}$
implies the following simple transformations for the combinations $U\tau^3
U^\dagger$ and $(D_{\mu}U)\tau^3U^\dagger$
\begin{eqnarray}
U\tau^3 U^\dagger & \mapsto & G_{L}\; U \tau^3 U^\dagger\; G_{L}^\dagger \\
(D_{\mu}U)\tau^3U^\dagger & \mapsto & G_{L}\; (D_{\mu}U)\tau^3U^\dagger \;
G_{L}^\dagger
\end{eqnarray}
Keeping all these relations in mind, we simply write down all the
possibilities for $\bar\psi\Gamma\psi$ and find the list of operators (\ref
{2.6b}--\ref{2.6d}). It is worth mentioning that there appears to be another
family of four operators in which the $U$ matrices also occur within a
trace: $\bar\psi \Gamma \psi\;\tr \Gamma^{\prime}$. One can check, however,
that these are not independent. More precisely
\begin{eqnarray}
\ri\bar q_L\ga^\mu q_L\;\tr (D_\mu\U)\tau^3\Ud&=&\Ap[3,4] \\
\ri\bar q_L\ga^\mu\U\tau^3\Ud q_L\;\tr (D_\mu\U)\tau^3\Ud&=&-\Ap[1,4]+%
\Ap[4,4] \\
\ri\bar q_R\ga^\mu q_R\;\tr (D_\mu\U)\tau^3\Ud&=&\Ap[5,4] \\
\ri\bar q_R\ga^\mu \tau^3 q_R\;\tr (D_\mu\U)\tau^3\Ud&=&\Ap[2,4]+\Ap[6,4]
\end{eqnarray}

Note that $\Ap[7,4]$ (as well as $\Ap[,R]{}^{\prime}$ discussed above) can
be reduced by equations of motion to operators of lower dimension which do
not contribute to the physical processes we are interested in. We have
checked that its contribution indeed drops from the relevant $S$-matrix
elements.

\section{Feynman rules}

\label{app-3.6b}

We write the effective $d=4$ lagrangian as
\begin{equation}
\Lc_{\rm eff}=\de^{\prime}\Ap[,R]{}^{\prime}+\sum_{k=1}^7\de_{k}\Ap[k,4]
\end{equation}
where $\de_{k}$ are real coefficients that we have to determine through the
matching. We need to match the effective theory described by $\Lc_{\rm eff}$ to
both, the MSM and the underlying theory parametrized by the four-fermion
operators. It has proven more convenient to work with the physical fields $%
W^\pm$, $Z$ and $\ga$ in the former case whereas the use of the lagrangian
fields $W^1$, $W^2$, $W^3$ and $B$ is clearly more straightforward for the
latter. Thus, we give the Feynman rules in terms of both the physical and
unphysical basis.

\setlength{\unitlength}{3mm}

\begin{eqnarray}
\begin{picture}(15,7)(-2,0) \put(0,1.5){\makebox(0,0)[c]{$d $}}
\put(11,1.5){\makebox(0,0)[c]{$\bar d$}} \put(0,0){\line(1,0){10}}
\put(2.5,0){\vector(1,0){0.1}} \put(7.5,0){\vector(1,0){0.1}}
\put(6,5){\makebox(0,0)[l]{$Z_\mu$}} \multiput(5,0.25)(0,1){5}{\zvs}
\end{picture}\kern-0.5cm &=& {\frac{\ri e}{2 s_W c_W}}\ga_\mu\left\{ \Half%
\left(-\de_1+\de_2+\de_4+\de_6 \right)-\de_3-\de_5\right.\nn \\
&&\phantom{{\ri e\over2 s_Wc_w}}\left. -\left( 1-{\frac{2}{3}} s^2_W \right)%
\de_7+{\frac{1}{3}} s^2_W\; \de^{\prime}\right\} \nn \\
&+&{\frac{\ri e}{2 s_W c_W}}\ga_\mu\ga_5\left\{ \Half\left(\de_1+\de_2-\de_4+%
\de_6 \right)+\de_3-\de_5\right.\nn \\
&&\phantom{{\ri e\over2 s_W c_W}}\left. +\left( 1-{\frac{2}{3}}s^2_W \right)%
\de_7+{\frac{1}{3}} s^2_W\; \de^{\prime}\right\}  \label{newN1d1} \\
\begin{picture}(15,7)(-2,0) \put(0,1.5){\makebox(0,0)[c]{$u $}}
\put(11,1.5){\makebox(0,0)[c]{$\bar u$}} \put(0,0){\line(1,0){10}}
\put(2.5,0){\vector(1,0){0.1}} \put(7.5,0){\vector(1,0){0.1}}
\put(6,5){\makebox(0,0)[l]{$Z_\mu$}} \multiput(5,0.25)(0,1){5}{\zvs}
\end{picture}\kern-0.5cm &=& {\frac{\ri e}{2 s_W c_W}}\ga_\mu\left\{ \Half%
\left(\de_1-\de_2-\de_4-\de_6 \right)-\de_3-\de_5\right.\nn \\
&&\phantom{{\ri e\over2 s_W c_W}}-\left. \left( 1-{\frac{4}{3}} s^2_W \right)%
\de_7+{\frac{2}{3}} s^2_W\; \de^{\prime}\right\} \nn \\
&+&{\frac{\ri e}{2 s_W c_W}}\ga_\mu\ga_5\left\{ \Half\left(-\de_1-\de_2+\de%
_4-\de_6 \right)+\de_3-\de_5\right.\nn \\
&&\phantom{{\ri e\over2 s_W c_W}}+\left. \left( 1-{\frac{4}{3}} s_W^2 \right)%
\de_7+{\frac{2}{3}}s_W^2\; \de^{\prime}\right\}  \label{newN1d2} \\
\begin{picture}(15,7)(-2,0) \put(0,1.5){\makebox(0,0)[c]{$d $}}
\put(11,1.5){\makebox(0,0)[c]{$\bar d$}} \put(0,0){\line(1,0){10}}
\put(2.5,0){\vector(1,0){0.1}} \put(7.5,0){\vector(1,0){0.1}}
\put(6,5){\makebox(0,0)[l]{$A_\mu$}} \multiput(5,0.25)(0,1){5}{\zvs}
\end{picture}\kern-0.5cm &=& -\ri e {\frac{1}{3}}\ga_\mu\left(\de_7+\Half \de%
^{\prime}\right) +\ri e{\frac{1}{3}}\ga_\mu\ga_5\left( \de_7-\Half\de%
^{\prime}\right)  \label{newN1d3} \\
\begin{picture}(15,7)(-2,0) \put(0,1.5){\makebox(0,0)[c]{$u $}}
\put(11,1.5){\makebox(0,0)[c]{$\bar u$}} \put(0,0){\line(1,0){10}}
\put(2.5,0){\vector(1,0){0.1}} \put(7.5,0){\vector(1,0){0.1}}
\put(6,5){\makebox(0,0)[l]{$A_\mu$}} \multiput(5,0.25)(0,1){5}{\zvs}
\end{picture}\kern-0.5cm &=& -\ri e {\frac{2}{3}}\ga_\mu\left(\de_7+\Half \de%
^{\prime}\right) +\ri e{\frac{2}{3}}\ga_\mu\ga_5\left( \de_7-\Half\de%
^{\prime}\right)  \label{newN1d4} \\
\begin{picture}(15,7)(-2,0) \put(0,1.5){\makebox(0,0)[c]{$d $}}
\put(11,1.5){\makebox(0,0)[c]{$\bar u$}} \put(0,0){\line(1,0){10}}
\put(2.5,0){\vector(1,0){0.1}} \put(7.5,0){\vector(1,0){0.1}}
\put(6,5){\makebox(0,0)[l]{$W^+_\mu$}} \multiput(5,0.25)(0,1){5}{\zvs}
\end{picture}\kern-0.5cm &=& -\ri e {\frac{1}{2\sqrt{2} s_W}}\ga_\mu\left(\de%
_1+\de_4-\de_2-\de_6\right)\nn \\
&+&\ri e {\frac{1}{2\sqrt{2} s_W}}\ga_\mu\ga_5\left(\de_1+\de_4+\de_2+\de_6
\right)  \label{newN1d5}
\end{eqnarray}

The operators $\Ap[7,4]$ and $\Ap[,4]{}^{\prime}$ contribute to two-point
function. The relevant Feynman rules are \setlength{\unitlength}{15mm}
\begin{eqnarray}
\begin{picture}(3,1)(-0.4,0) \put(0,0.375){\makebox(0,0)[c]{$u$}}
\put(2,0.375){\makebox(0,0)[c]{$\bar u $}}
\put(1.0,0){\makebox(0,0)[c]{$\times$}} \put(0,0){\line(1,0){2}}
\put(0.5,0){\vector(1,0){0.02}} \put(1.5,0){\vector(1,0){0.02}} \end{picture}%
\kern-0.5cm &=& \ri(\de_7+\half \de') \ps +\ri(-\de_7+\half \de') \ps \ga_5
\label{new-1} \\
\begin{picture}(3,1)(-0.4,0) \put(0,0.375){\makebox(0,0)[c]{$d$}}
\put(2,0.375){\makebox(0,0)[c]{$\bar d $}}
\put(1.0,0){\makebox(0,0)[c]{$\times$}} \put(0,0){\line(1,0){2}}
\put(0.5,0){\vector(1,0){0.02}} \put(1.5,0){\vector(1,0){0.02}} \end{picture}%
\kern-0.5cm &=& \ri(-\de_7-\half \de') \ps +\ri(\de_7-\half \de') \ps \ga_5
\label{new-2}
\end{eqnarray}

Rather than giving the actual Feynman rules in the unphysical basis, we
collect the various tensor structures that can result from the calculation
of the relevant diagrams in table~\ref{table-1}.
\begin{table}[h]
\centering
\[
\begin{array}{l||c|c|c|c|c|c|}
\mbox{Tensor structure} & \de_1 & \de_2 & \de_3 & \de_4 & \de_5 & \de_6 \\
\hline
\ri\bar q_L\,g[\Wi] q_L & 1 &  &  & 1 &  &  \\
\ri\bar q_L\,\tau^3[g\Ws^3-g^{\prime}\Bs] q_L & 1 &  &  & -1 &  &  \\
\ri\bar q_L\,[g\Ws^3-g^{\prime}\Bs] q_L &  &  & -1 &  &  &  \\
\ri\bar q_R\,g[\Wi] q_R &  & -1 &  &  &  & 1 \\
\ri\bar q_R\,\tau^3[g\Ws^3-g^{\prime}\Bs] q_R &  & -1 &  &  &  & -1 \\
\ri\bar q_R\,[g\Ws^3-g^{\prime}\Bs] q_R &  &  &  &  & -1 &
\end{array}
\]
\caption{Various structures appearing in the matching of the vertex and the
corresponding contributions to $\de_{1}$, \dots, $\de_{6}$.}
\label{table-1}
\end{table}
We include only those that can be matched to insertions of the operators $%
\Ap[1,4],\dots, \Ap[6,4]$ (the contributions to $\Ap[7,4]$ and $\Ap[,4]{}%
^{\prime}$ can be determined from the matching of the two-point functions).
The corresponding contributions of these structures to $\de_{1},\dots\de_{6}$
are also given in table~\ref{table-1}. Once $\de_{7}$ has been replaced by
its value, obtained in the matching of the two-point functions, only the
listed structures can show up in the matching of the vertex, otherwise the $%
SU(2)\times U(1)$ symmetry would not be preserved.

\section{Four-fermion operators}

The complete list of four-fermion operators relevant for the present
discussion is in tables~\ref{table-2} and~\ref{table-3} in
section~\ref{new-phys}. It is also explained in sec.~\ref{new-phys} the convenience
of fierzing the operators in the last seven rows of
table~\ref{table-2} in order to write them in the form ${\bf
J}\cdot{\bf j}$. Here we just give the list that comes out naturally
from our analysis, tables~\ref{table-2} and~\ref{table-3},
without further physical interpretation.
The list is given for fermions belonging to the
representation ${\bf 3}$ of $SU(3)_{c}$ (techniquarks). By using Fierz
transformations one can easily find out relations among some of these
operators when the fermions are colour singlet (technileptons),
which is telling us that some of these operators are not independent
in this case. A list of independent operators for technileptons is also
given in sec.~\ref{new-phys}.

Let us outline the procedure we have followed to obtain this basis in
the (more involved) case of coloured fermions.

There are only two colour singlet structures one can build out of
four fermions, namely 
\begin{eqnarray}
(\bar \psi \psi)(\bar \psi^{\prime}\psi^{\prime})&\equiv& \bar \psi_\al
\psi_\al\; \bar \psi^{\prime}_\be \psi^{\prime}_\be  \label{26.5d} \\
(\bar \psi \vec\la \psi)\cdot(\bar \psi^{\prime}\vec\la \psi^{\prime})
&\equiv& \bar \psi_\al (\vec\la)_{\al\be} \psi_\be\;\cdot\; \bar
\psi^{\prime}_\ga (\vec\la)_{\ga\de} \psi^{\prime}_\de,  \label{26.5e}
\end{eqnarray}
where, $\psi$ stands for any field belonging to the
representation ${\bf 3}$ of $SU(3)_{c}$ ($\psi$ will be either $q$ or $Q$);
$\al$, $\be$, \dots, are colour indices; and the primes (${}'$) remind
us that $\psi$ and $\bar\psi$ carry same additional indices (Dirac,
$SU(2)$, \dots).

Next we clasify the Dirac structures. Since $\psi$ is either
$\psi_{L}$ [it
belongs to the representation $(\half,0)$ of the Lorentz group] or $\psi_{R}$
[representation $(0,\half)$], we have five sets of fields to analyse, namely
\begin{eqnarray}
&&\{\bar\psi_{L},\psi_{L},\bar\psi^{\prime}_{L},\psi^{\prime}_{L}\},\quad
[R\leftrightarrow L];\qquad
\{\bar\psi_{L},\psi_{L},\bar\psi_{R},\psi_{R}\};
\label{26.5b} \\
&&\{\bar\psi_{L},\psi_{R},\bar\psi^{\prime}_{L},\psi^{\prime}_{R}\},\qquad
[R\leftrightarrow L].  \label{26.5c}
\end{eqnarray}
There is only an independent scalar we can build with each of the three sets
in~(\ref{26.5b}). Our choice is
\begin{eqnarray}
& &\bar \psi_L\ga^\mu \psi_L\; \bar \psi^{\prime}_L \ga_\mu \psi^{\prime}_L
,\qquad [R\leftrightarrow L];  \label{28.5-59} \\
& &\bar \psi_L\ga^\mu \psi_L\; \bar \psi_R \ga_\mu \psi_R.  \label{28.5-60}
\end{eqnarray}
where the prime is not necessary in the second equation because $R$ and
$L$ suffice to remind us that the two $\psi$ and $\bar\psi$ may carry
different ($SU(2)$, technicolour, \dots) indices.
There appear to be four other independent scalar operators: $\bar \psi_L\ga%
^\mu \psi^{\prime}_L\; \bar \psi^{\prime}_L \ga_\mu \psi_L$, $%
[R\leftrightarrow L]$; $\bar \psi_L \psi_R\; \bar \psi_R \psi_L$; and $\bar
\psi_L \si^{\mu\nu} \psi_R\;\bar \psi_R \si_{\mu\nu} \psi_L$. However, Fierz
symmetry implies that the first three are not independent, and the fourth
one
vanishes, as can be also seen using the identity $2\ri\si^{\mu\nu}\ga^5=\ep%
^{\mu\nu\rho\la} \si_{\rho\la} $. For each of the two operators in~(\ref
{26.5c}), two independent scalars can be constructed. Our choice is
\begin{eqnarray}
& &\bar \psi_L \psi_R\; \bar \psi^{\prime}_L \psi^{\prime}_R,\qquad
[R\leftrightarrow L];  \label{28.5-68} \\
& &\bar \psi_L \psi^{\prime}_R\; \bar \psi^{\prime}_L \psi_R,\qquad
[R\leftrightarrow L].  \label{28.5-69}
\end{eqnarray}
Again, there appear to be four other scalar operators: $\bar \psi_L\si%
^{\mu\nu} \psi_R \; \bar \psi^{\prime}_L\si_{\mu\nu} \psi^{\prime}_R$, $%
[R\leftrightarrow L]$;\break $\bar \psi_L\si^{\mu\nu} \psi^{\prime}_R \;
\bar \psi^{\prime}_L\si_{\mu\nu} \psi_R$, $[R\leftrightarrow L]$; which,
nevertheless, can be shown not to be independent but related to (\ref
{28.5-68}) and (\ref{28.5-69}) by Fierz symmetry. To summarize, the
independent scalar structures are (\ref{28.5-59}), (\ref{28.5-60}), (\ref
{28.5-68}) and (\ref{28.5-69}).

Next, we combine the colour and the Dirac structures. We do this for the
different cases (\ref{28.5-59}) to (\ref{28.5-69}) separately. For operators
of the form (\ref{28.5-59}), we have the two obvious possibilities
(Hereafter, colour and Dirac indices will be implicit)
\begin{eqnarray}
& & (\bar \psi_L \ga^\mu \psi_L) (\bar \psi^{\prime}_L \ga_\mu
\psi^{\prime}_L), \qquad [R\leftrightarrow L];  \label{26.5f} \\
& & (\bar \psi_L \ga^\mu \psi^{\prime}_L) (\bar \psi^{\prime}_L \ga_\mu
\psi_L), \qquad [R\leftrightarrow L];  \label{26.5g}
\end{eqnarray}
where fields in parenthesis have their colour indices contracted as in~(\ref
{26.5d}) and~(\ref{26.5e}). Note that the operator $(\bar \psi_L \ga^\mu \vec%
\la \psi_L)\cdot (\bar \psi^{\prime}_L \ga_\mu\vec\la \psi^{\prime}_L)$, or
its $R$ version, is not independent (recall that
$(\vec\la)_{\al\be}\cdot (\vec\la)_{\ga\de}=2\de_{\al\de}\de_{\be\ga}
-2/3\;\de_{\al\be}\de_{\ga\de}$). For operators of the form (\ref{28.5-60}%
), we take
\begin{eqnarray}
& & (\bar \psi_L \ga^\mu \psi_L) (\bar \psi_R \ga_\mu \psi_R)  \label{C1} \\
& & (\bar \psi_L \ga^\mu \vec\la \psi_L)\cdot (\bar \psi_R \ga_\mu\vec\la %
\psi_R)  \label{C2}
\end{eqnarray}
Finally, for operators of the form (\ref{28.5-68}) and~(\ref{28.5-69}), our
choice is
\begin{eqnarray}
& & (\bar \psi_L \psi_R)(\bar \psi^{\prime}_L \psi^{\prime}_R),\quad
[R\leftrightarrow L];\quad (\bar \psi_L \vec\la \psi_R)\cdot(\bar
\psi^{\prime}_L \vec \la \psi^{\prime}_R),\quad [R\leftrightarrow L];
\label{26.5i} \\
& & (\bar \psi_L \psi^{\prime}_R)(\bar \psi^{\prime}_L \psi_R),\quad
[R\leftrightarrow L];\quad (\bar \psi_L \vec\la \psi^{\prime}_R)\cdot(\bar
\psi^{\prime}_L \vec \la \psi_R),\quad [R\leftrightarrow L].  \label{26.5k}
\end{eqnarray}
All them are independent unless further symmetries [e.g., $SU(2)_L\times
SU(2)_R$] are introduced.

To introduce the $SU(2)_L\times SU(2)_R$ symmetry one just assigns $SU(2)$
indices ($i$, $j$, $k$, \dots) to each of the fields
in~(\ref{26.5f}--\ref{26.5k}). We can drop the primes hereafter since
there is no other symmetry left but technicolour which for the present
analysis is trivial (recall that we are only interested in four
fermion operators of the form $Q\bar Qq\bar q$, thus technicolour
indices must necessarily be matched in the obvious way: $Q^A\bar Q^A q\bar q$).
For each of the operators in~(%
\ref{26.5f}) and~(\ref{26.5g}), there are two independent ways of
constructing $SU(2)_L\times SU(2)_R$ invariants. Only two of the four
resulting operators turn out to be independent (actually, the other two are
exactly equal to the first ones). The independent operators are chosen to be
\begin{eqnarray}
& &(\bar \psi_L^i \ga^\mu \psi_L^i)(\bar \psi^j_L \ga_\mu \psi^j_L)\equiv
(\bar \psi_L \ga^\mu \psi_L)(\bar \psi_L \ga_\mu \psi_L), \quad
[R\leftrightarrow L];  \label{D1} \\
& &(\bar \psi_L^i \ga^\mu \psi_L^j)(\bar \psi^j_L \ga_\mu \psi^i_L), \quad
[R\leftrightarrow L];  \label{D2}
\end{eqnarray}
For each of the operators in~(\ref{C1}--\ref{26.5k}), the same
straightforward group analysis shows that there is only one way to construct
a $SU(2)_L\times SU(2)_R$ invariant. Discarding the redundant operators and
imposing hermiticity and $CP$ invariance
one finally has, in addition to the operators~(\ref{D1}%
) and~(\ref{D2}), those listed below (from now on, we understand that fields
in parenthesis have their Dirac, colour and also flavour indices
contracted as in~(\ref{D1}))
\begin{eqnarray}
&&(\bar \psi_L\ga^\mu \psi_L)(\bar \psi_R\ga_\mu \psi_R),  \label{1.6a} \\
&&(\bar \psi_L\ga^\mu\vec\la \psi_L)\cdot (\bar \psi_R\ga_\mu\vec\la \psi_R),
\label{1.6b} \\
&&(\bar \psi_L^i \psi_R^j)(\bar \psi_L^k \psi_R^l) \ep_{ik}\ep_{jl}+ (\bar
\psi_R^i \psi_L^j)(\bar \psi_R^k \psi_L^l) \ep_{ik}\ep_{jl},  \label{1.6c} \\
&&(\bar \psi_L^i\vec\la \psi_R^j)\cdot(\bar \psi_L^k\vec\la \psi_R^l) \ep%
_{ik}\ep_{jl} +(\bar \psi_R^i\vec\la \psi_L^j)\cdot(\bar \psi_R^k\vec\la %
\psi_L^l) \ep_{ik}\ep_{jl}.  \label{1.6d}
\end{eqnarray}

We are now in a position to obtain very easily the custodially preserving
operators of tables~\ref{table-2} and~\ref{table-3} 
We simply
replace $\psi$ by $q$ and $Q$ (a pair of each: a field and its conjugate) in
all possible independent ways.

To break the custodial symmetry we simply insert $\tau^3$ matrices in the $R$%
-sector of the custodially preserving operators we have just obtain
(left columns of
tables~\ref{table-2} and~\ref{table-3}).
However, not all the operators obtained this
way are independent since one can prove the following relations
\begin{eqnarray}
(\bar q_R^i\ga^\mu Q_R^j)(\bar Q_R^j\ga_\mu [\tau^3 q_R]^i)
&=&(\bar q_R\ga^\mu\tau^3 Q_R)(\bar Q_R\ga_\mu q_R)
+(\bar q_R\ga^\mu Q_R)(\bar Q_R\ga_\mu \tau^3 q_R)\nn\\
& &\qquad-
(\bar q_R^i\ga^\mu [\tau^3 Q_R]^j)(\bar Q_R^j\ga_\mu q_R^i)\\
(\bar q_R^i\ga^\mu [\tau^3 Q_R]^j)(\bar Q_R^j\ga_\mu [\tau^3 q_R]^i)&=&
(\bar q_R\ga^\mu Q_R)(\bar Q_R\ga_\mu q_R)
+(\bar q_R\ga^\mu\tau^3  Q_R)(\bar Q_R\ga_\mu \tau^3 q_R)\nn
\\ &&\qquad-
(\bar q_R^i\ga^\mu  Q_R^j)(\bar Q_R^j\ga_\mu q_R^i)\\
(\bar q_R^i\ga^\mu [\tau^3 q_R]^j)(\bar Q_R^j\ga_\mu [\tau^3 Q_R]^i)&=&
(\bar q_R\ga^\mu q_R)(\bar Q_R \ga_\mu Q_R)+
(\bar q_R\ga^\mu\tau^3 q_R)(\bar Q_R \ga_\mu\tau^3 Q_R)\nn\\
&&\qquad-
(\bar q_R^i\ga^\mu q_R^j)(\bar Q_R^j\ga_\mu Q_R^i)\\
(\bar q_R^i\ga^\mu [\tau^3 q_R]^j)(\bar Q_R^j\ga_\mu Q_R^i)\qquad
\phantom{+}\nn\\
+
(\bar q_R^i\ga^\mu  q_R^j)(\bar Q_R^j\ga_\mu [\tau^3 Q_R]^i)&=&
(\bar q_R\ga^\mu q_R)(\bar Q_R\ga^\mu \tau^3 Q_R)\nn\\
&&\qquad+
(\bar q_R\ga^\mu \tau^3 q_R)(\bar Q_R\ga^\mu  Q_R).
\end{eqnarray}

Our final choice of custodially breaking operators is the one in the right
columns of tables~\ref{table-2} and~\ref{table-3}.

%

\section{Renormalization of the matter sector}

Although most of the material in this section is standard, it is convenient
to collect some of the important expressions, as the
renormalization of the fermion fields is somewhat involved and also to set
up
the notation. Let us introduce three wave-function renormalization constants
for the fermion fields \bea
\pmatrix{u \cr d}_L&\to& Z_L^{1/2}\pmatrix{u \cr d}_L, \nn \\u_R&\to&
(Z_R^u)^{1/2} u_R,\nn \\d_R&\to& (Z_R^d)^{1/2} d_R.\label{newglo4} \eea
where $u$ ($d$) stands for the field of the up-type (down-type) fermion. We
write \beq
Z_i=1+\de Z_i \eeq
We also renormalise the fermion masses according to \bea
m_f&\to& m_f+\de m_f, \eea
where $f=u,\; d$. These substitutions generate the counterterms needed to
cancel the UV divergencies. The corresponding Feynman rules are \bea
\setlength{\unitlength}{3mm}
\begin{picture}(15,4)(-2,0)
\put(0,1.5){\makebox(0,0)[c]{$q$}}
\put(11,1.5){\makebox(0,0)[c]{$\bar q $}}
\put(5.12,0){\makebox(0,0)[c]{$\times$}}
\put(0,0){\line(1,0){10}}
\put(2.5,0){\vector(1,0){0.1}}
\put(7.5,0){\vector(1,0){0.1}}
\end{picture}
=&\phantom{-}& \ri\de Z_V^f p\slash -\ri \de Z_A^f p\slash \ga_5 -\ri\left({%
\frac{\de m_f}{m_f}}+\de Z_V^f\right)\label{17.6a}\\\setlength{%
\unitlength}{3mm}
\begin{picture}(15,7)(-2,0)
\put(0,1.5){\makebox(0,0)[c]{$q $}}
\put(11,1.5){\makebox(0,0)[c]{$\bar q$}}
\put(5.25,0){\makebox(0,0)[c]{$\times$}}
\put(0,0){\line(1,0){10}}
\put(2.5,0){\vector(1,0){0.1}}
\put(7.5,0){\vector(1,0){0.1}}
\put(6,5){\makebox(0,0)[l]{$Z_\mu$}}
\multiput(5,0.25)(0,1){5}{\zvs}
\end{picture}
=&-& \ri e\ga_\mu (v_f-a_f\,\ga_5) (\dZZp-\dZZ)\nn\\&-&\ri e\ga_\mu\, Q_f\, (%
\dZZgap-\dZZga)\nn\\&-&\ri e\ga_\mu (v_f\, \de Z_V^f+a_f\,\de Z_A^f)\nn\\&+&%
\ri e\ga_\mu \ga_5(v_f\, \de Z_A^f+a_f\,\de Z_V^f)\label{17.6h}\\%
\setlength{\unitlength}{3mm}
\begin{picture}(15,7)(-2,0)
\put(0,1.5){\makebox(0,0)[c]{$q $}}
\put(11,1.5){\makebox(0,0)[c]{$\bar q$}}
\put(5.25,0){\makebox(0,0)[c]{$\times$}}
\put(0,0){\line(1,0){10}}
\put(2.5,0){\vector(1,0){0.1}}
\put(7.5,0){\vector(1,0){0.1}}
\put(6,5){\makebox(0,0)[l]{$A_\mu$}}
\multiput(5,0.25)(0,1){5}{\zvs}
\end{picture}
=&-&\ri e \ga_\mu \, Q_f (\dZgap-\dZga+\de Z_V^f-\de Z_A^f\,\ga_5)\nn\\&-&%
\ri e\ga_\mu (v_f- a_f\,\ga_5) (\dZZgap-\dZZga) \\\setlength{%
\unitlength}{3mm}
\begin{picture}(15,7)(-2,0)
\put(0,1.5){\makebox(0,0)[c]{$d $}}
\put(11,1.5){\makebox(0,0)[c]{$\bar u$}}
\put(5.25,0){\makebox(0,0)[c]{$\times$}}
\put(0,0){\line(1,0){10}}
\put(2.5,0){\vector(1,0){0.1}}
\put(7.5,0){\vector(1,0){0.1}}
\put(6,5){\makebox(0,0)[l]{$W^+_\mu$}}
\multiput(5,0.25)(0,1){5}{\zvs}
\end{picture}
=&-&\ri {\frac{e}{2\sqrt{2} s_W}} \ga_\mu (1-\ga_5) \, (\dZWp-\dZW+\de Z_L)
\label{17.6i} \eea
Here we have introduced the notation \beq
\de Z_L=\de Z_V^{u,d}+\de Z_A^{u,d},\quad \de Z_R^{u,d}=\de Z_V^{u,d}-\de %
Z_A^{u,d}, \eeq
and \beq
v_f={\frac{I^3_f-2 Q_f s_W^2}{2 s_W c_W}},\qquad a_f={\frac{I^3_f}{2 s_W c_W}%
}. \eeq
Note that the Feynman rules for the vertices contain additional
renormalization constants which should be familiar from the oblique
corrections.

The fermion self-energies can be decomposed as \beq
\Si^f(p)=p\slash\, \Si_V^f(p^2)+p\slash\ga_5\, \Si_A^f(p^2)+ m\, \Si%
_S^f(p^2) \eeq

By adding the conterterms one obtains de renormalised self-energies, which
admit the same decomposition. One has \bea
\hat\Si_V^f(p^2)&=& \Si_V^f(p^2)-\de Z_V^f,\label{17.6d}\\\hat\Si%
_A^f(p^2)&=& \Si_A^f(p^2)+\de Z_A^f,\label{17.6e}\\\hat\Si_S^f(p^2)&=& \Si%
_S^f(p^2)+{\frac{\de m_{f}}{m_{f}}}+\de Z_V^f, \label{17.6f} \eea
where the hat denotes renormalised quantities. The on-shell renormalization
conditions amount to \bea
{\frac{\de m_{u,d}}{m_{u,d}}}&=&-\Si_V^{u,d}(m^2_{u,d})-\Si%
_S^{u,d}(m^2_{u,d})\label{9.6a}\\\de
Z_V^d&=&\Si_V^d(m^2_d)+2m^2_d[\Si^d_V{}'(m^2_d)+\Si^d_S{}'(m^2_d)] \label
{17.6b}\\\de Z_A^{u,d}&=&-\Si_A^{u,d}(m^2_{u,d})\label{17.6c} \eea
where $\Si{}^{\prime}(m^2)=[\p \Si(p^2) /\p p^2]_{p^2=m^2}$. Eq.~(\ref{9.6a}%
) guarantees that $m_{u}$, $m_{d}$ are the physical fermion masses. The
other two equations, come from requiring that the residue of the down-type
fermion be unity. One cannot simultaneously impose this condition to both
up- and down-type fermions. Actually, one can easily work out the residue of
the up-type fermions which turns out to be $1+\de_\res$ with \beq
\de \res=\hat\Si_V^u(m^2_u)+2 m^2_u\left[\hat\Si^u_V{}'(m^2_u)+ \hat\Si%
^u_S{}'(m^2_u) \right]. \label{17.6j} \eeq

\section{Effective lagrangian coefficients}

In this appendix we shall provide the general expressions for the
coefficients $a_i$ and $\delta _i$ in theories of the type we have been
considering. The results are for the usual
representations of $SU(2)\times SU(3)_c$. Extension to
other representations is possible using the prescriptions
listed in section 8.
\begin{eqnarray}
a_0 &=&\frac{n_{TC} n_D }{64\pi ^2M_Z^2s_W^2}\left( \frac{%
m_2^2+m_1^2}2+
\frac{m_1^2m_2^2\ln \frac{m_1^2}{m_2^2}}{m_2^2-m_1^2}\right)
+\frac{1}{16\pi^2}\frac{3}{8}(\frac{1}{\hat\epsilon}-
\log\frac{\Lambda^2}{\mu^2}),
\\
a_1 &=&-\frac{n_{TC} n_D }{96\pi ^2}+\frac{n_{TC}\left(
n_Q-3n_L\right) }{3\times 96\pi ^2}\ln \frac{m_1^2}{m_2^2}
+\frac{1}{16\pi^2}\frac{1}{12}(\frac{1}{\hat\epsilon}-
\log\frac{\Lambda^2}{\mu^2}),
\\
a_8 &=&-\frac{n_{TC}\left( n_c+1\right) }{96\pi ^2}\frac 1{\left(
m_2^2-m_1^2\right) ^2}\left\{ \frac 53m_1^4-\frac{22}3m_2^2m_1^2+\frac
53m_2^4\right.  \\
&&+\left. \left( m_2^4-4m_2^2m_1^2+m_1^4\right) \frac{m_2^2+m_1^2}{%
m_2^2-m_1^2}\ln \frac{m_1^2}{m_2^2}\right\} ,
\end{eqnarray}
where $n_{TC}$ the number of technicolors (taken equal to 2 in all numerical
discussions), $n_D$ is the number of technidoublets. It is interesting to
note that all effective lagrangian coefficients (except for $a_1$) depend
on $n_D$ and are independent of the actual hypercharge (or charge)
assignment. $n_Q$ and $n_L$ are the actual number of techniquarks
and technileptons. In the one-generation model $n_Q=3$, $n_L=1$ and,
consequently,  $n_D=4$. Furthermore in this model $a_1$ is
mass independent.
For simplicity we have written $m_1$ for the dynamically generated
mass of the $u$-type technifermion and $m_2$ for the one of the $d$-type,
and assumed that they are the same for all doublets. This is of course
quite questionable as a large splitting between the
technielectron and the technineutrino seems more likely and they should
not necessarily coincide with techniquark masses, but
the appropriate expressions can be easily inferred from the above formulae
anyway.
\begin{eqnarray}
\de_{1} & = & {n_{D} n_{TC} G^2 \over16\pi^2M^2} a_{\vec L^2}
\left\{ {m_1^2+m_2^2\over2}-m_1^2\left(
1+{m_1^2\over m_1^2-m_2^2}\right)\log{m_{1}^2\over
M^2 }\right.\nn\\
        & & - m_2^2\left.\left(
1+{m_2^2\over m_2^2-m_1^2} \right)\log{m_{2}^2\over
M^2} \right\}
     \label{aa1}  \\
\de_{2} & = & {n_{D} n_{TC} G^2 \over16\pi^2M^2}
     \left\{\left(a_{LR_{3}}-a_{RR_{3}}\right)A_{-}+a_{R_{3}^2}
     A_{+}+a_{\vec R^2} B_{+}\right\}
     \label{aa2}  \\
     \de_{3} & = & {n_{D} n_{TC} G^2 \over16\pi^2M^2}
     \left\{\left(a_{L^2}-a_{RL}\right)A_{-}+a_{R_{3}L}
     A_{+}\right\}
     \label{aa3}  \\
\de_{4} & = & {n_{D} n_{TC} G^2 \over16\pi^2M^2}\; a_{\vec L^2}
\left\{
{m_1^2+m_2^2\over2}+m_1^2\left(
       1-{m_1^2\over m_1^2-m_2^2}
                                \right)\log{m_{1}^2\over
                    M^2 }\right.\nn\\
& & + m_2^2\left.\left( 1-{m_2^2\over m_2^2-m_1^2} \right)\log{m_{2}^2\over
M^2}\right\}
     \label{aa4}  \\
     \de_{5} & = & {n_{D} n_{TC} G^2 \over16\pi^2M^2}
     \left\{\left(a_{LR}-a_{R^2}\right)A_{-}+a_{R_{3}R}
     A_{+}\right\}
     \label{aa5}  \\
     \de_{6} & = & {n_{D} n_{TC} G^2 \over16\pi^2M^2}
     \left\{\left(a_{LR_{3}}-a_{RR_{3}}\right)A_{-}+a_{R_{3}^2}
     A_{+}+a_{\vec R^2} B_{-}\right\}
     \label{aa6}  \\
     \de_{7} & = & 0
     \label{aa7}
\end{eqnarray}
where
\begin{eqnarray}
     A_{\pm} & = & \mp
     m_{1}^2\log{m_{1}^2\over M^2}-m_{2}^2\log{m_{2}^2\over M^2}
     \label{bb1}  \\
     B_{\pm} & = & \pm 2m_{1}m_{2}-
     m_{1}^2\left(1\pm{2m_{1}m_{2}\over m_{1}^2-m_{2}^2}\right)
     \log{m_{1}^2\over M^2}\nn\\
&-& m_{2}^2 \left(1\pm{2m_{2}m_{1}\over m_{2}^2-m_{1}^2}\right)
\log{m_{2}^2\over M^2}.         \label{bb2}
\end{eqnarray}
We have not bothered to write the chiral divergences counterterms
in the above expressions. They are identical to those of section 8.
Although we have written the full expressions obtained using
chiral quark model methods, one should be well aware of
the approximations made in the text.

\vfill
\eject

\end{document}